\documentclass[aps,twocolumn,showpacs,preprintnumbers,prb,amsmath,amssymb,amsfonts,superscriptaddress,floatfix]{revtex4-1}

\usepackage[T1]{fontenc}
\usepackage[latin1]{inputenc}
\usepackage{graphics}
\usepackage{color}

\usepackage{amssymb}
\usepackage{amsmath}
\usepackage{overpic}
\usepackage{epstopdf}
\usepackage{bm}
\usepackage{graphicx}
\usepackage{subfigure}

\newcommand{\be}{\begin{equation}}
\newcommand{\ee}{\end{equation}}
\newcommand{\bea}{\begin{eqnarray}}
\newcommand{\eea}{\end{eqnarray}}

\newcommand{\lasco}{La$_{2-x}$Sr$_{x}$CuO$_{4}$}
\newcommand{\ybco}{YBa$_{2}$Cu$_{3}$O$_{7-x}$}
\newcommand{\bscco}{Bi$_{2}$Sr$_{2}$CaCu$_{2}$O$_{8+x}$}

\newcommand{\angstrom}{\textup{\AA}}
\def\m{\mu}
\def\l{\lambda}

\def\s{\sigma}

\def\D{\Delta}
\def\ra{\rightarrow}

\def\pll{\parallel}

\def\Ra{\Rightarrow}

\def\tbkt{$T_{BKT}$}

\def\lb{\label}
\def\pref#1{(\ref{#1})}

\newcount\bozza \bozza=1
\ifnum\bozza=1
\newdimen\shift \shift=-2truecm
\def\lb#1{%
{\label{#1}\rlap{\kern\shift{$\scriptstyle#1$}}}}
\else\def\lb#1{\label{#1}} \fi

\begin{document}

\title{Effective 2D thickness for the Berezinskii-Kosterlitz-Thouless-like transition in a highly underdoped \lasco}

\author{P.~G.~Baity}
\affiliation{National High Magnetic Field Laboratory, Florida State University, Tallahassee, Florida 32310, USA}
\affiliation{Department of Physics, Florida State University, Tallahassee, Florida 32306, USA}
\author{Xiaoyan Shi}
\altaffiliation{Present address: Department of Physics, University of Texas at Dallas, Richardson, TX 75080, USA}
\affiliation{National High Magnetic Field Laboratory, Florida State University, Tallahassee, Florida 32310, USA}
\affiliation{Department of Physics, Florida State University, Tallahassee, Florida 32306, USA}
\author{Zhenzhong Shi}
\affiliation{National High Magnetic Field Laboratory, Florida State University, Tallahassee, Florida 32310, USA}
\author{L.~Benfatto} 
\affiliation{ISC-CNR and Dept. of Physics, Sapienza University of Rome,
  P.le A. Moro 5, 00185, Rome, Italy}
\author{Dragana Popovi\'c}
\email{dragana@magnet.fsu.edu}
\affiliation{National High Magnetic Field Laboratory, Florida State University, Tallahassee, Florida 32310, USA}
\affiliation{Department of Physics, Florida State University, Tallahassee, Florida 32306, USA}

\date{\today}

\begin{abstract}
The nature of the superconducting transition in highly underdoped thick films of \lasco\, ($x=0.07$ and 0.08) has been investigated using the in-plane transport measurements.  The contribution of superconducting fluctuations to the conductivity in zero magnetic field, or paraconductivity, was determined from the magnetoresistance measured in fields applied perpendicular to the CuO$_2$ planes.  Both the temperature dependence of the paraconductivity above the transition and the nonlinear current-voltage ($I-V$) characteristics measured across it, exhibit the main signatures of the Berezinskii-Kosterlitz-Thouless (BKT) transition.  The quantitative comparison of the superfluid stiffness, extracted from the $I-V$ data, with the renormalization-group results for the BKT theory, reveals a large value of the vortex-core energy.  This finding is confirmed by the analysis of the paraconductivity obtained using different methods.  The results strongly suggest that the characteristic energy scale controlling the BKT behavior in this layered system corresponds to the superfluid stiffness of a few layers.
\end{abstract}

% fluctuations in sc, hole-doped cuprates, disorder effects in Tc

\maketitle

\section{Introduction}
\label{intro}

One of the most  intriguing phenomena in condensed matter systems is the occurrence of the so-called  Berezinskii-Kosterlitz-Thouless\cite{bkt,bkt1,bkt2} (BKT) transition in two-dimensional (2D) superfluid systems. The main ingredients of the BKT physics were described originally within the context of the  two-dimensional XY model, which is an effective model for the collective phase of the superfluid order parameter.\cite{review_minnaghen,jose_book} Here logarithmically interacting vortex-like topological excitations drive the transition from the superfluid state, where they are bound together in vortex-antivortex (V-AV) pairs, to the metallic one, where single vortex excitations proliferate. This mechanism leads in principle to several peculiar signatures in the physical observables, such as the universal and discontinuous jump\cite{nelson_prl77} of the superfluid density at \tbkt, the observation of which in $^4$He films\cite{helium4} was the first experimental proof of the existence of a BKT transition.  Afterwards, interest in BKT physics was triggered mainly by the possibility to observe it in superconducting (SC) systems that can be considered to be in the 2D limit. On very general grounds, this occurs for systems with low superfluid stiffness $J_s$, defined as the energy scale associated to the areal density of superfluid electrons:
\be
\label{jbkt}
J_s=\frac{\hbar^2 n_s d_{BKT}}{4m}=\frac{\hbar^2 c^2}{16\pi e^2}\frac{d_{BKT}}{\lambda^2},
\ee
where $n_s,m$ denote the superfluid density and mass of the carriers, respectively, $\lambda$ is the magnetic penetration depth and $d_{BKT}$ denotes a transverse length scale over which the system can be seen as effectively 2D.  The possibility to see BKT physics is connected to a low value of $d_{BKT}/\lambda^2$:
indeed, despite the presence of screening supercurrents, the interaction between vortices remains logarithmic when the Pearl screening length $\Lambda=2\lambda^2/d_{BKT}$ overcomes the system size.\cite{Pearl}  In addition, since the distance between \tbkt \, and the ordinary BCS temperature $T_c$ scales as $(T_c-T_{BKT})/T_c\propto T_{BKT}/J_s$, a clear BKT regime can only be identified when $J_s$ gets reduced.  In films of conventional superconductors, these conditions are usually realized when the film thickness $d$ is reduced.  In those cases, by identifying $d_{BKT}$ with $d$,  typical BKT signatures have been observed\cite{fiory_prb83,lemberger_prl00,armitage_prb07,armitage_prb11,kamlapure_apl10,mondal_bkt_prl11,goldman_prl12,yazdani_prl13} by means of different experimental probes. The universal jump of the superfluid density has been seen either via direct measurements of the inverse penetration depth\cite{fiory_prb83, lemberger_prl00,armitage_prb07,armitage_prb11,kamlapure_apl10,mondal_bkt_prl11,yazdani_prl13} or via a discontinuous jump of the  exponent of the nonlinear $I-V$ characteristics.\cite{fiory_prb83}  At the same time, the vortex proliferation above \tbkt\, has been identified\cite{fiory_prb83,mondal_bkt_prl11,goldman_prl12} from an exponential divergence of the correlation length above \tbkt, which leads to a peculiar paraconductivity above the transition.\cite{review_minnaghen,benfatto_review13}  

An alternative route for the observation of BKT physics is presented by {\em bulk} layered systems, in which the magnetic-field distribution of a vortex differs drastically from the monopole-like Pearl solution in uniform films:\cite{review_minnaghen,blatter_review} the presence of other superconducting layers squeezes the field of a pancake vortex into a narrow strip of size $\lambda$ along the $c$ axis.  This in turn implies that the logarithmic dependence of the interaction potential between two vortices placed in the same layer persists up to all length scales, as in a neutral superfluid, making in principle the stack of uncoupled layers the best possible system to observe a true BKT transition, with the 2D unit in Eq.\ \pref{jbkt} corresponding to each isolated plane. In the presence of Josephson coupling between layers, the upper cut-off for the logarithmic interaction between vortices becomes\cite{review_minnaghen,blatter_review} $\Lambda_J\simeq \xi_0/\sqrt{J_\perp/J_\pll}$, where $\xi_0$ is the zero-temperature in-plane  coherence length, and $J_{\pll,\perp}$ are the in-plane and out-of-plane superfluid stiffness, respectively.  If the interlayer coupling is weak, \textit{i.e.}  $J_\perp/J_\pll\ll 1$,  this length scale is large enough to allow for a BKT-like description of the vortex-antivortex interaction, independent of the film thickness $d$. In practice, even if the finite-size effect due to $\Lambda_J$ leads to a rounding of the discontinuous jump in $J_s$, the analysis of anisotropic 3D $XY$-like model\cite{shenoy_prl94,friesen_prb95,pierson_prb95,olsson_prb91,benfatto_mu_prl07,sondhi_prb09} shows that the unbinding of vortex-antivortex pairs in each plane is still the mechanism driving the transition, in analogy with the purely 2D case.  Therefore, in a weakly coupled, layered superconductor, one expects to observe a BKT-like transition at a 3D transition temperature that is slightly higher than the BKT transition temperature of a single layer of an equivalent uncoupled system.

Such a description is expected to be appropriate for underdoped samples of cuprate superconductors,  which are highly anisotropic, layered materials.  Here one also finds that the superfluid stiffness is suppressed by the proximity to the Mott insulator,\cite{emery_95,review_lee} making the separation between \tbkt\, and $T_c$ large, while avoiding the additional consequences of an increase of the disorder level, as it occurs in films of conventional superconductors  
when the thickness is reduced.  According to this argument,  in bulk samples of underdoped cuprates one should be able to identify BKT signatures assuming that the fundamental 2D unit is represented by isolated CuO$_2$ layers, \textit{i.e.} the transverse length scale $d_{BKT}$ in Eq.\ \pref{jbkt} would coincide with the interlayer distance $d_c$, as pointed out in the seminal work by Emery and Kivelson.\cite{emery_95}  However, it has been recently shown\cite{benfatto_mu_prl07} that this picture is somehow too simplified, since one should also account for the nontrivial role of the vortex-core energy $\mu$, which is the energetic cost needed to create the vortex at the smallest length scale $\xi_0$. Indeed, even if the layers are weakly coupled, what matters for the vortex proliferation is the competition at large distances between the effective vortex fugacity and the effective Josephson coupling.  As a consequence, when $\mu$ is large, the Josephson coupling between layers can prevent the vortex unbinding, moving the BKT transition away from the value expected for each isolated layer, resulting in an effective dimension $d_{BKT}$ larger than $d_c$.

So far, the experimental situation in cuprate superconductors has been controversial. For example, the direct measurements of the inverse penetration depth  have shown that, in the \ybco \, family, no BKT jump is observed even in strongly-underdoped thick films\cite{lemberger_prl05,broun_prl07} or crystals.\cite{hardy_prl94} A BKT-like superfluid-density jump is only seen in few-unit-cell thick films of \ybco\, (Ref.~\onlinecite{lemberger_natphys07}) or \bscco(Ref.~\onlinecite{lemberger_prb12}), but even in this case, as the samples get underdoped, the effective $d_{BKT}$ seems to cross over to the sample thickness and the superfluid-density jump gets smeared out. While this can be explained indeed by an increase of the vortex-core energy with underdoping,\cite{benfatto_bilayer_prb08} one should notice that the simultaneous appearance of an anomalously large dissipative response suggests that spurious finite-frequency effects can also be present, as emphasized recently  in the analysis of thin films of NbN.\cite{ganguli_prb15} These spurious effects are instead absent in the dc measurements of the $I-V$ exponent that suggested a BKT-like jump very near $T_c$ in cuprate 
samples.\cite{forro_prb88,yeh_prb89,norton_prb93,vidal_cm13,bozovic_prb15}  However, this measurement allows one to extract directly the effective 2D areal stiffness \pref{jbkt}, \textit{i.e.} the combination $d_{BKT}/\lambda^2$, so $d_{BKT}$ can be determined only if $\l$ is known by measurements in similar samples. Finally, the analysis of the paraconductivity, \textit{i.e.} of the SC fluctuations above $T_c$, also raises some questions on the occurrence or not of a BKT transition. Indeed, on one hand, the SC fluctuations have been proved to have a strong 2D character in several cuprate families (\textit{e.g.} Refs.~\onlinecite{Balestrino_prb89, Balestrino_prb92, caprara_prb05,leridon_prb07,rullier_prb11}) with the typical 2D unit being identified as the distance between the CuO$_2$ layers $d_c$.  On the other hand, these are ordinary Gaussian (amplitude and phase) fluctuations, with a BKT regime that, if present, is restricted to a small range of temperatures near $T_c$ in the most underdoped samples.\cite{rullier_prb11,vidal_cm13}

In the present work, we address the issue of the identification of the scale $d_{BKT}$ in cuprate superconductors by making a simultaneous analysis of the BKT signatures both below and above \tbkt\, in two highly underdoped samples of \lasco. We first extract the paraconductivity above $T_{BKT}$ (Sec.~\ref{exp:para}), and then determine the temperature dependence of the anomalous 2D exponent of the $I-V$ characteristics across it (Sec.~\ref{exp:IV}). In 
Sec.~\ref{th-J}, the direct comparison of the experimental $I-V$ data with the renormalization-group results for the BKT theory allows us to extract a large value of the vortex-core energy $\mu$, consistent with that obtained from the analysis of paraconductivity in Sec.~\ref{exp:para}.  According to earlier theoretical work,\cite{benfatto_mu_prl07,benfatto_bilayer_prb08} the large value of $\mu$ obtained in our study corresponds to $d_{BKT}\simeq (2-3)\, d_c$.
Furthermore, this value of the vortex-core energy can be used to reduce considerably the fitting parameters in the well-known Halperin-Nelson formula\cite{HN} for the paraconductivity above $T_{BKT}$, spanning both the BKT and Aslamazov-Larkin\cite{AL1,AL2,Varlamov_book} (AL) regimes of the SC fluctuations.  This analysis (Sec.~\ref{th-para}) confirms that the effective length scale $d_{BKT}$ is a few times larger than $d_c$, in agreement with the expectation\cite{benfatto_mu_prl07,benfatto_bilayer_prb08} for a layered weakly-coupled system with a large vortex-core energy. Our study clarifies how different transverse length scales enter in the analysis of the SC fluctuations above and below $T_{BKT}$, solving the apparent contradiction between previous measurements. 

\section{Experiments}
\subsection{Samples and measurement techniques}

The  samples were \lasco\, (LSCO) films with the nominal doping $x=0.07$ and $x=0.08$.  They were patterned into standard Hall bars with the length $L=2.0$~mm and the width $W=0.3$~mm; the distance between voltage contacts was 1.01~mm.  The films were 75 unit cells (150 CuO$_2$ layers) thick ($d\approx 1000$~\AA) and grown by molecular beam epitaxy.  The films and samples were described in detail elsewhere.\cite{Shi2013}  The samples become superconductors below the temperature $T_{R=0}(x)$, defined as the temperature at which the in-plane resistance $R$ becomes zero.  The measured $T_{R=0}$ were $(3.9\pm 0.1)$~K and $(9.7\pm 0.3)$~K for samples $x=0.07$ and 0.08, respectively.

The in-plane sample resistance and magnetoresistance were measured in $^3$He cryostats (base $T\approx 0.25$~K) with a standard four-probe ac method ($\sim$13-16~Hz) in the Ohmic regime, using either SR7265 lock-in amplifiers or a LR-700 resistance bridge.  The magnetic fields $H$ up to 18~T were applied perpendicular to CuO$_2$ planes ($H\parallel c$ axis) and swept at constant temperatures.  The sweep rates of 0.02-1~T/min were low enough to avoid the heating of the sample from eddy currents.

The current-voltage ($I$-$V$) measurements were carried out at constant temperatures ($T$) in $H=0$ using $^3$He and variable-temperature insert (base $T\approx 1.3$~K) cryostats.  DC square pulses provided by a Keithley 6221 current source were applied to the samples, while a Keithley 2182A nanovoltmeter measured the voltage response. Each data point on the $I$-$V$ curve was found by averaging measurements with positive and negative pulse polarities. Such a four-point dc method\cite{Keithley} avoids possible effects of parasitic capacitances (\textit{e.g.} from the sample contacts) 
and obviates Joule heating, while retaining the increased sensitivity of a finite-frequency technique and eliminating the effects of thermal electromotive forces.  Current excitations between 50~nA and 1~mA were typically used, depending on the film doping and temperature.

The addition of current noise to a device with an intrinsic nonlinear behavior can create an Ohmic response at low currents\cite{Sullivan2004} and, in particular, it can create Ohmic behavior even below \tbkt.  Therefore, for the $I$-$V$ measurements, filtering was provided at room temperature  by a 1.75~nF low-pass $\pi$ filter in series with a 1~k$\Omega$ resistor on each lead to the sample.  The $\pi$ filters and the resistors were encased in a shielded box attached to the top of the cryostat probe.  This filter box provided a 5~dB (60~dB) noise reduction at 10~MHz (1~GHz), which enabled the observation of nonlinear $I$-$V$ behavior at low excitations amid masking current noise.  

\subsection{High-field magnetoresistance measurements and superconducting fluctuations}
\label{exp:para}

By approaching the superconducting transition from above, it is in principle possible to identify the BKT transition from the temperature dependence of the
contribution of superconducting fluctuations (SCFs) to conductivity (or ``paraconductivity''), $\Delta\sigma_{SCF}(T)=\rho(T)^{-1}-\rho_{n}(T)^{-1}$, where $\rho (T)$ and $\rho_n(T)$ are the measured and normal-state resistivity, respectively.  In cuprates, the determination of $\rho_n(T)$ has been somewhat ambiguous and controversial (see, \textit{e.g.}, Ref.~\onlinecite{rullier_prb11} and references therein).  We emphasize, however, that the precise determination of (finite) $\rho_n$ is not crucial for the extraction of $\Delta\sigma_{SCF}$ in the regime of interest, very near the BKT transition where the contribution of SCFs diverges (Eq.\ \pref{para} below).  On the other hand, it may introduce considerable errors into the values of $\Delta\sigma_{SCF}$ far from it.\cite{rullier_prb11}  This issue is demonstrated and discussed further in Sec.~\ref{th-para}.

In this Section, we adopt a method that uses transverse ($H\parallel c$) magnetoresistance measurements to determine the extent of SCFs.  In particular, above a sufficiently high magnetic field $H_{c}'(T)$, SCFs are completely suppressed (\textit{i.e.} they become unobservable in the experiment) and the normal state is fully restored.  In the normal state, the magnetoresistance of cuprates increases as $H^2$ at low fields\cite{Lacerda1994, Harris1995, Kimura1996, Ando2002, Vanacken2004, Cooper2009, Shi2013, rullier_prl07, rullier_prb11, Rourke2011, Shi2014, Chan2014} ($\omega_{c}\tau\ll 1$, where $\omega_c$ is the cyclotron frequency and $\tau$ is the scattering time), similar to the classical orbital effect in conventional metals:\cite{Pippard}
\begin{equation}
\label{mr-orb}
\frac{\rho_{n}(H) -\rho_{n}(0)}{\rho_{n}(0)}= (\omega_{c}\tau)^2 \propto H^2,
\end{equation}
Therefore, the values of $H_{c}'$ can be found from the downward deviations from such quadratic dependence that arise from SCFs when $H<H_{c}'$.\cite{Shi2013, rullier_prl07, rullier_prb11,Rourke2011,Shi2014}  The SCF contribution to the conductivity can be determined then as $\Delta\sigma_{SCF}(T,H)=\rho(T,H)^{-1}-\rho_{n}(T,H)^{-1}$, where $\rho (T,H)$ is the measured resistivity and $\rho_n(T,H)$ is obtained by extrapolating the region of $H^2$ magnetoresistance observed at high enough $H$ and $T$.  The advantages of this method\cite{rullier_prl07,rullier_prb11} over some of the earlier ones (\textit{e.g.} Refs.~\onlinecite{caprara_prb05,lang_prb95}) are that it does not rely on any assumptions about the $T$ dependence of $\rho_{n}$,  
and it makes it possible to determine both the paraconductivity $\Delta\sigma_{SCF}(T,H=0)$ and the SCF contribution to conductivity in the presence of magnetic field.

%
%%%%%%%%%%%% Figure 1 %%%%%%%%%%%%
%
\begin{figure}
\centering
\includegraphics[width=8cm]{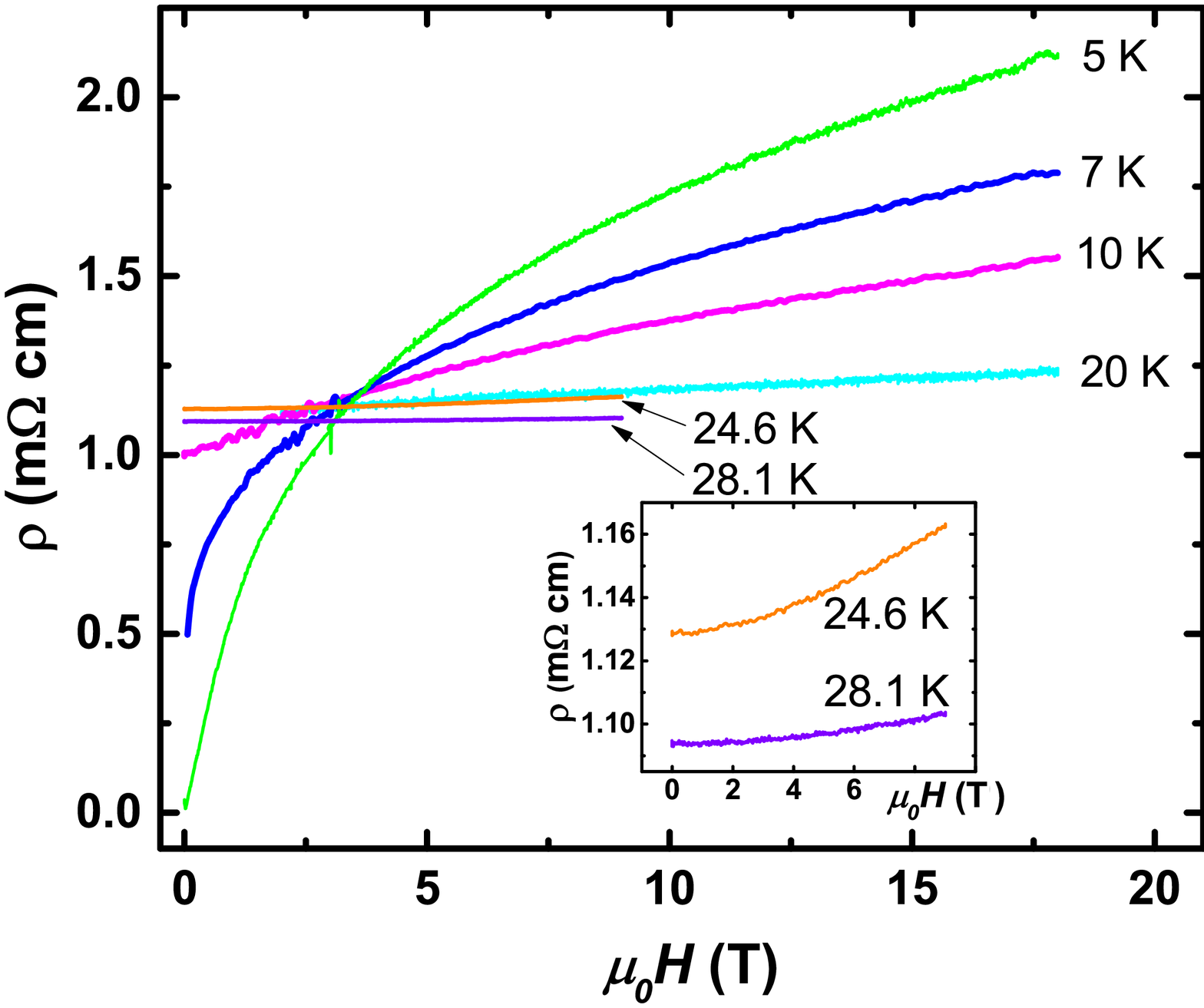}\llap{\parbox[b]{0.9in}{\textbf{(a)}\\\rule{0ex}{0.6in}}}
\includegraphics[width=8cm]{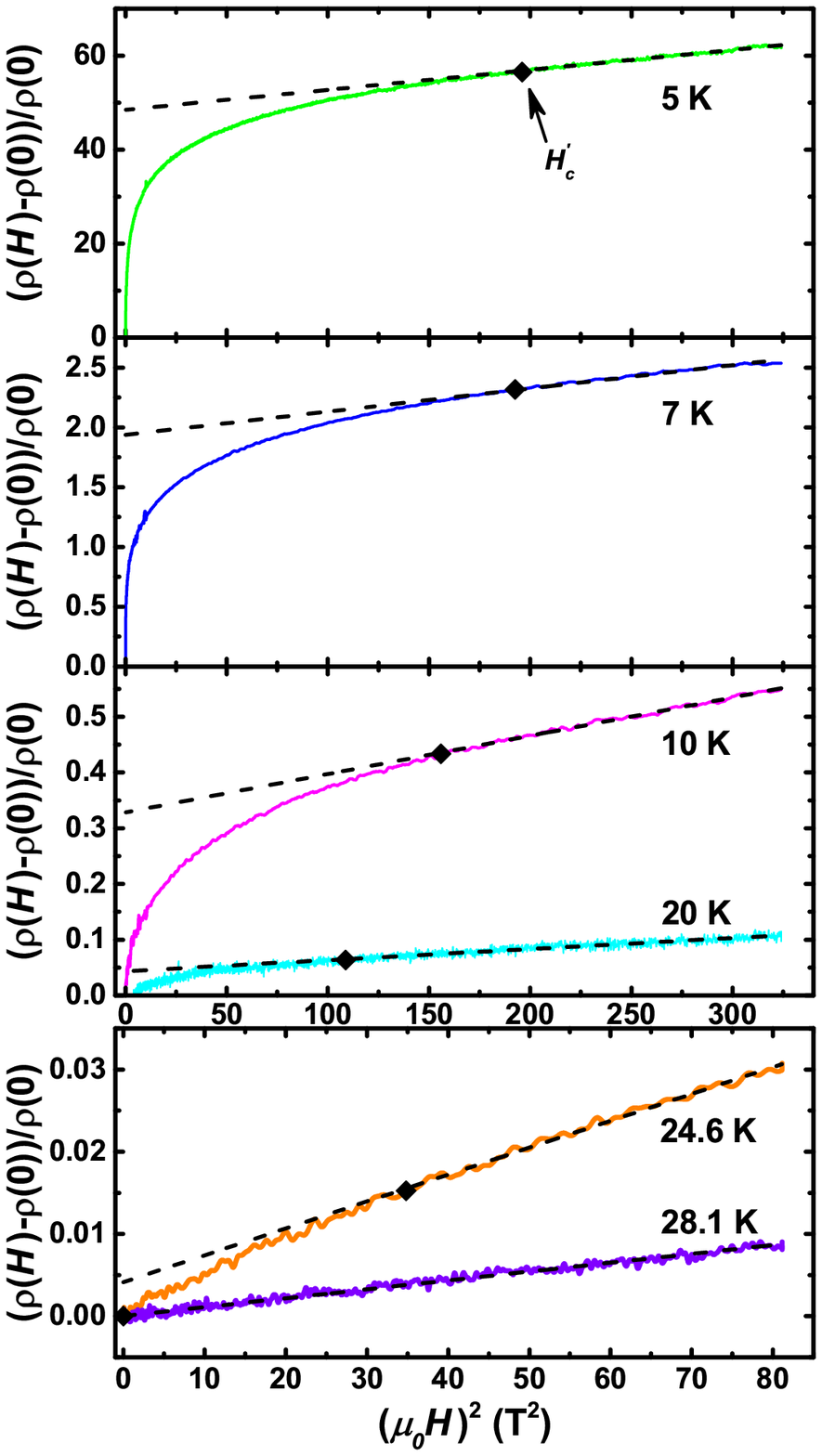}\llap{\parbox[b]{0.9in}{\textbf{(b)}\\\rule{0ex}{5.4in}}}
\caption{(Color online) $x=0.07$ LSCO film.  (a) Resistivity \textit{vs.} transverse magnetic field ($H\parallel c$) up to 18~T for different $T$, as shown.  The highest $T$ data are also shown in the inset for clarity.  (b) Magnetoresistance data from (a) plotted vs $H^2$.   Symbols (black diamonds) show $H_{c}'(T)$, the fields above which SCFs are fully suppressed and the $H^2$ dependence of the normal-state resistivity $\rho_{n}$ is observed.  Dashed lines are linear fits representing the contributions from normal state transport, \textit{i.e.} they correspond to $[\rho(H)-\rho(0)]/\rho(0)=[\rho_n(0)-\rho(0)]/\rho(0)+[\rho_n(0)/\rho(0)]a_{trans}H^2$.}\label{fig:MR}
\end{figure}
%%%%%%%%%%%%%%%%%%%%%%%%%%%%
%

Figure~\ref{fig:MR}(a)  shows representative $\rho(H)$ curves ($H\parallel c$) obtained on the $x=0.07$ LSCO sample.  The condition for the weak-field limit is satisfied in the entire regime of interest as $\omega_{c}\tau\approx0.5$ at 18~T and  5~K, where it reaches its maximum value. 
By tracking the gradual evolution of the magnetoresistance curves measured at different $T$ [Fig.~\ref{fig:MR}(b)], from the high-$T$ region where the $H^2$ dependence is unambiguous, to lower $T$ where SCFs are more pronounced, we were able to determine the values of the onset fields $H_{c}'(T)$ (see Appendix~\ref{app-MR} for a more detailed discussion).  

\begin{figure}
\centering
\includegraphics[width=8cm]{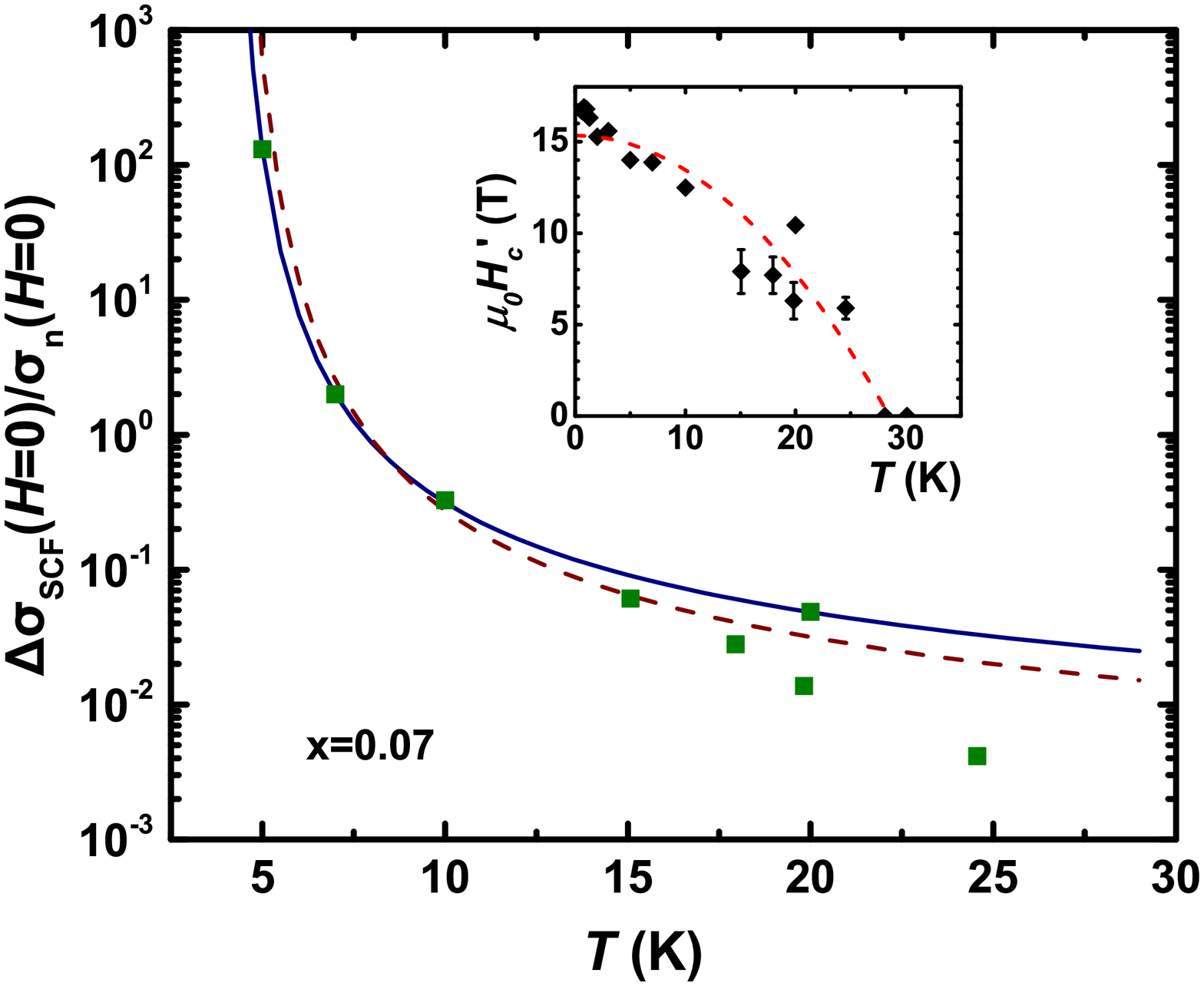}\llap{\parbox[b]{0.9in}{\textbf{(a)}\\\rule{0ex}{2.0in}}}
\includegraphics[width=8cm]{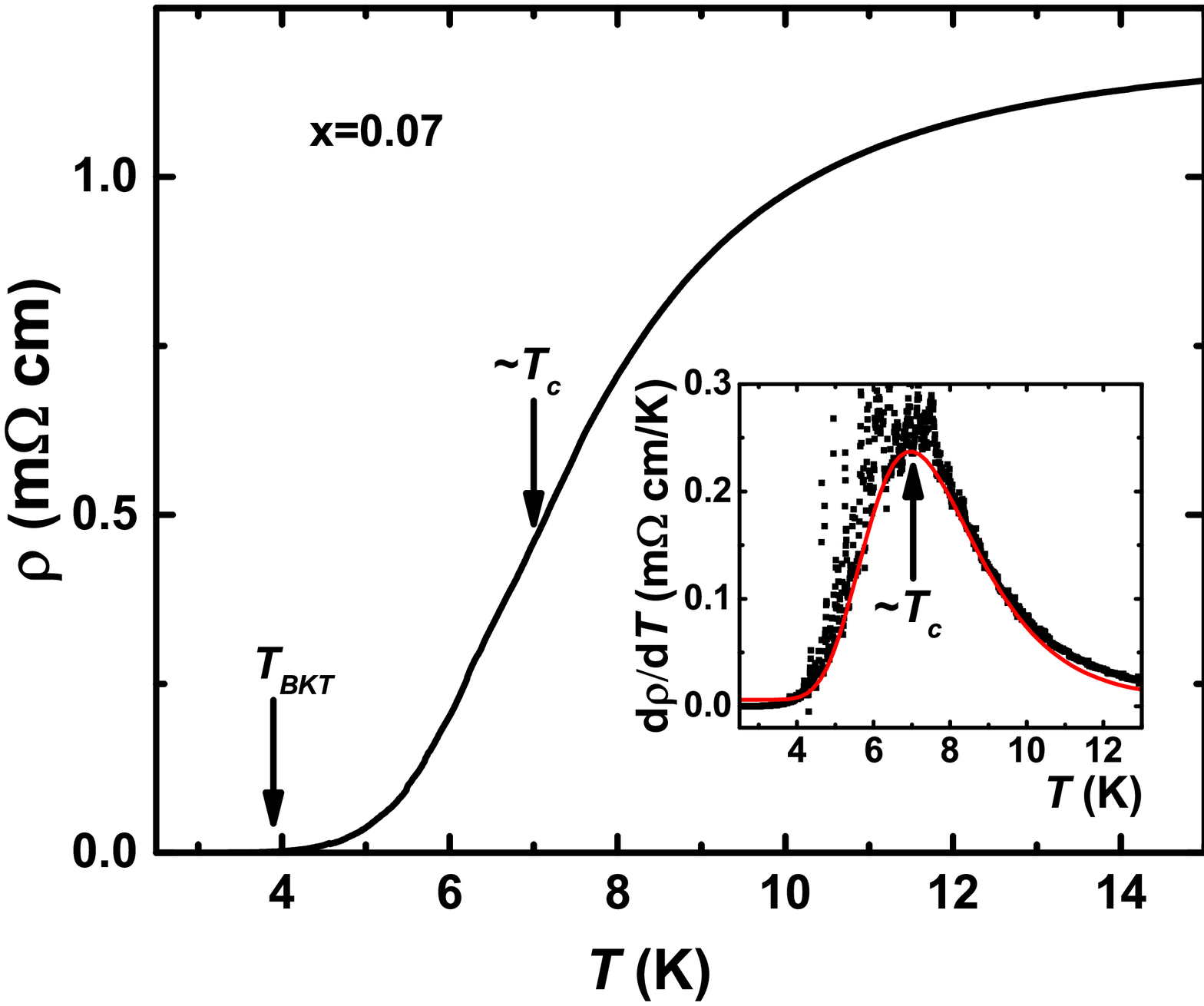}\llap{\parbox[b]{0.9in}{\textbf{(b)}\\\rule{0ex}{2.0in}}}
\caption{(Color online) $x=0.07$ LSCO film.  (a) Symbols show $\Delta\sigma_{SCF}(T,H=0)/\sigma_{n}(H=0)$ vs $T$, as determined from Fig.~\ref{fig:MR}.  The solid line is a fit to Eq.~\pref{HN-LB} with $T_{BKT}=3.8$~K and fitting parameters $A=17.9$, $b=2.9$; the dashed line corresponds to $T_{BKT}=4.0$~K, $A=29$, $b=3.3$.  Inset: The onset field $H_{c}'(T)$ below which SCFs become observable.  The dashed line is a fit $H_{c}'=H_{c}'(0)[1-(T/T_2)^2]$, with $\mu_{0}H_{c}'(0)=15$~T and $T_2=29$~K.  (b)  $\rho$ vs $T$ in $H=0$.  Arrows point at \tbkt$=T_{R=0}$ and $T_c$; $T_c$ was estimated as shown in the inset.
Inset: The temperature at the inflection point of the $\rho(T)$ curve, where $d\rho/dT$ has a maximum, is taken as an estimate of $T_c$ 
in the calculation of $\alpha$ from Eq.~\pref{best}.}\label{fig:SCF}
\end{figure}

Figure~\ref{fig:SCF}(a) inset shows $H_{c}'(T)$, determined from Fig.~\ref{fig:MR}(b) for the $x=0.07$ sample and fitted by a simple quadratic expression $H_{c}'(T)=H_{c}'(0)[1-(T/T_2)^2]$, similar to earlier studies.\cite{rullier_prl07, rullier_prb11, Rourke2011, Shi2013,Shi2014}  In zero field, SCFs become observable below $T_2=29$~K.                                                                                                                                                                                                                                                                                                                                            
%
%%%%%%%%% Figure 2 %%%%%%%%%%%%%%%
%
%%%%%%%%%%%%%%%%%%%%%%%%%%%
%
In Sec.~\ref{th-para}, we show explicitly that the exact determination of $\rho_n$, and thus the determinations of $H_{c}'(T)$ and $T_2$, does not affect our conclusions.  Hereafter we focus only on the zero-field behavior.

Figure~\ref{fig:SCF}(a) shows that $\Delta\sigma_{SCF}(H=0)/\sigma_n(H=0)$, where $\sigma_n=1/\rho_n$, increases by several orders of magnitude as temperature is reduced towards $T_{R=0}\approx 4$~K, reminiscent of the exponential divergence expected at the BKT transition.  Indeed, in 2D the paraconductivity can always be expressed as
\be
\label{para}
\Delta\sigma_{SCF}/\sigma_n=\left[\frac{\xi(T)}{\xi_0}\right]^2,
\ee
where $\xi(T)$ is the SC  correlation length, whose temperature dependence depends on the nature of the SC fluctuations. The usual  Aslamazov-Larkin\cite{AL1,AL2,Varlamov_book}   (AL) paraconductivity describes the fluctuating Cooper pairs above the  mean-field temperature $T_c$, and leads to a power-law divergence of the coherence length  $\xi^{2}\sim(T-T_{c})^{-1}$. In contrast, within BKT theory, $\xi^2(T)\sim 1/n_F$ measures the inverse density $n_F$ of free vortices above $T_{BKT}$, and diverges exponentially as $T \ra$~\tbkt.  An interpolation formula between these two regimes was first proposed by Halperin and Nelson\cite{HN} 
\begin{equation}
\label{HN-LB}
\frac{\Delta\sigma_{SCF}}{\sigma_{n}}=\left(\frac{2}{A}\sinh \frac{b}{\sqrt t}\right)^2, \quad T \gtrsim T_{BKT},
\end{equation}
where $t=(T-T_{BKT})/T_{BKT}$, and $A$ and $b$ are numerical constants.  More recently, a renormalization-group (RG) study\cite{benfatto_inho_prb09}
 of the BKT transition showed that parameter $b$ is strictly connected to two relevant physical quantities:
\be
\label{best}
b\simeq 2\alpha \sqrt{t_c}, \quad \alpha=\mu/\mu_{XY},
\ee
where $t_c$ is the distance between the mean-field and BKT critical temperatures
\be
\label{red}
t_c\equiv \frac{T_c-T_{BKT}}{T_{BKT}},
\ee
while $\alpha$ is the vortex-core energy $\mu$ expressed in units of the conventional value $\mu_{XY}$ that it assumes in the XY model (see also Eq.\ \pref{def_muxy} below). According to Eq.\ \pref{para}, the exponential BKT behavior is limited to the range of temperatures $t\ll t_c$, while above it, one recovers the usual AL paraconductivity. 

The paraconductivity shown in Fig.~\ref{fig:SCF}(a) has been fitted to Eq.~\pref{HN-LB} by taking $T_{BKT}=T_{R=0}=(3.9\pm 0.1)$~K [Fig~\ref{fig:SCF}(b)].  Surprisingly, it is possible to get a good fit to the data even up to very high temperatures $\sim 20$~K with reasonable values of $A$ and $b$ [\textit{e.g.} dashed line in Fig.~\ref{fig:SCF}(a)].  However, within the error for $T_{BKT}$, the lower-$T$ data up to $\sim 10$~K are described better with the fitting parameters in the range $A=13-20$ and $b=2.5-3.0$ [\textit{e.g.} solid line with $A=17.9$ and $b=2.9$ in Fig.~\ref{fig:SCF}(a)].  Assuming that $T_c\sim 7$~K, \textit{i.e.} of the order of the temperature where $d\rho/dT$ has a maximum [Fig~\ref{fig:SCF}(b) inset], Eq.~\pref{best} then yields enhanced values of the vortex-core energy, $\mu/\mu_{XY}\simeq 1.4-1.7$, consistent with previous work.\cite{benfatto_mu_prl07,benfatto_bilayer_prb08}

The above analysis of the SCFs above a SC transition, which occurs at $T_{BKT}=T_{R=0}$, suggests the presence of a BKT fluctuation regime at  $T_{BKT}<T<T_c\sim 7$~K, followed by a crossover to the AL regime at $T>T_c$.  It is worth noting that the crossover to the AL regime gives some indication on the transverse length scale controlling the Gaussian fluctuations in the sample. Indeed, when $t\gg t_c$, Eq.\ \pref{HN-LB} reduces to
\be
\label{sbkt}
\frac{\D\s_{SCF}}{\s_n}\simeq \frac{4b^2}{A^2}\frac{T_{BKT}}{T-T_{BKT}} \simeq \kappa_{BKT}\frac{T_{c}}{T-T_c},
\ee
where, on the r.h.s.,  $\kappa_{BKT}\equiv 4b^2 T_{BKT}/(A^2T_c)$ and we replaced $T-T_{BKT}$ with $\simeq T-T_c$, which is correct when $T$ is sufficiently larger than $T_c$ so that the difference between $T_c$ and $T_{BKT}$ can be neglected. We note that, in films of conventional superconductors,\cite{mondal_bkt_prl11,yong_prb13} usually $t_c$ is at most of order 0.1, so the crossover from the pure BKT behavior to the AL one occurs for relatively small reduced temperatures $t$. In our samples, $t_c$ is as large as $0.7$, so the asymptotic AL behavior \pref{sbkt} is reached at higher temperatures. On the other hand, since the SCFs regime extends up to reduced temperatures  as large as $t\sim 3 - 4$ [Fig.~\ref{fig:SCF}(a)], there is still a large temperature regime where the approximation \pref{sbkt} is valid. This expression has to be compared with the usual AL formula\cite{AL1,AL2,Varlamov_book}  that gives
\be
\label{sal}
\frac{\D\s_{AL}}{\s_n}= \frac{\rho_n/d_{AL}}{16 R_c}  \frac{T_c}{T-T_{c}}\equiv \kappa_{AL}\frac{T_c}{T-T_{c}},
\ee
where $R_c=\hbar/e^2=4.1$ k$\Omega$.  By mapping the expressions \pref{sbkt} and \pref{sal}, we can see that the high-$T$ limit of the interpolating HN formula also fixes the prefactor $\kappa_{AL}$ that controls the strength of the AL fluctuations in the Gaussian regime at $T\geq T_c$.  The latter one depends in turn on the transverse length scale $d_{AL}$ that identifies the 2D unit for AL fluctuations [see Eq.\ \pref{sal}]. By using the estimates of $b, A$ given above, we obtain that $\kappa_{BKT}=4b^2 T_{BKT}/(A^2T_c)\simeq 0.1$. Thus, from the measured $\rho_n\simeq 1$ m$\Omega$ cm and by matching $\kappa_{BKT}$ and $\kappa_{AL}$, we conclude that $d_{AL}$ is of the same order as the interlayer distance $d_c$, in full agreement with previous work in the literature.\cite{caprara_prb05,leridon_prb07} In other words, as far as the Cooper-pair fluctuations are concerned, the fluctuation regime displays marked 2D character with decoupled layers, consistent with the standard expectation for a weakly-coupled layered superconductor.\cite{Varlamov_book}  On the other hand,  the BKT paraconductivity does not allow us to extract any precise information on the scale $d_{BKT}$ controlling the vortex physics below $T_{BKT}$. To address this issue, and to confirm the fit based on the paraconductivity data extracted from the high-field magnetoresistance measurements [Fig.~\ref{fig:SCF}(a)], we analyze the $I-V$ characteristics, whose behavior  is, in fact, one of the key signatures of the BKT transition.

\subsection{Current-voltage characteristics and superfluid stiffness}
\label{exp:IV}

The most famous hallmark of the BKT transition is observed by approaching \tbkt\, from below.  In particular, the superfluid stiffness $J_s$, defined in Eq.\ \pref{jbkt}, exhibits the so-called universal jump at the transition, \textit{i.e.}
\be
\label{jump}
J_s(T^-_{BKT})=\frac{2}{\pi} T_{BKT}, \quad J_s(T_{BKT}^+)=0.
\ee
Here the $T$ dependence of $J_s(T)$ includes both the quasiparticle excitations, which would drive $J_s$ continuously to zero at $T_c$,  and vortex-like phase fluctuations, which are instead responsible for the discontinuous jump \pref{jump}. The latter directly influences
the behavior of the exponent $a$ in the $I-V$ characteristics:
\be
\label{alpha}
V\propto I^{a(T)}, \quad a(T)=\frac{\pi J_s(T)}{T}+1.
\ee
The superlinear behavior in Eq.~\pref{alpha} is due to the ability of a sufficiently large current to unbind vortex-antivortex pairs.  From Eq.~\pref{jump}, it follows then that $a$ should jump from $a=3$ at $T=T^-_{BKT}$ to $a=1$ at $T=T^+_{BKT}$.  Below $T_{BKT}$, the exponent $a$ is expected to increase with decreasing $T$ since the superfluid density increases.

\begin{figure}
\centering
\includegraphics[width=8.5cm]{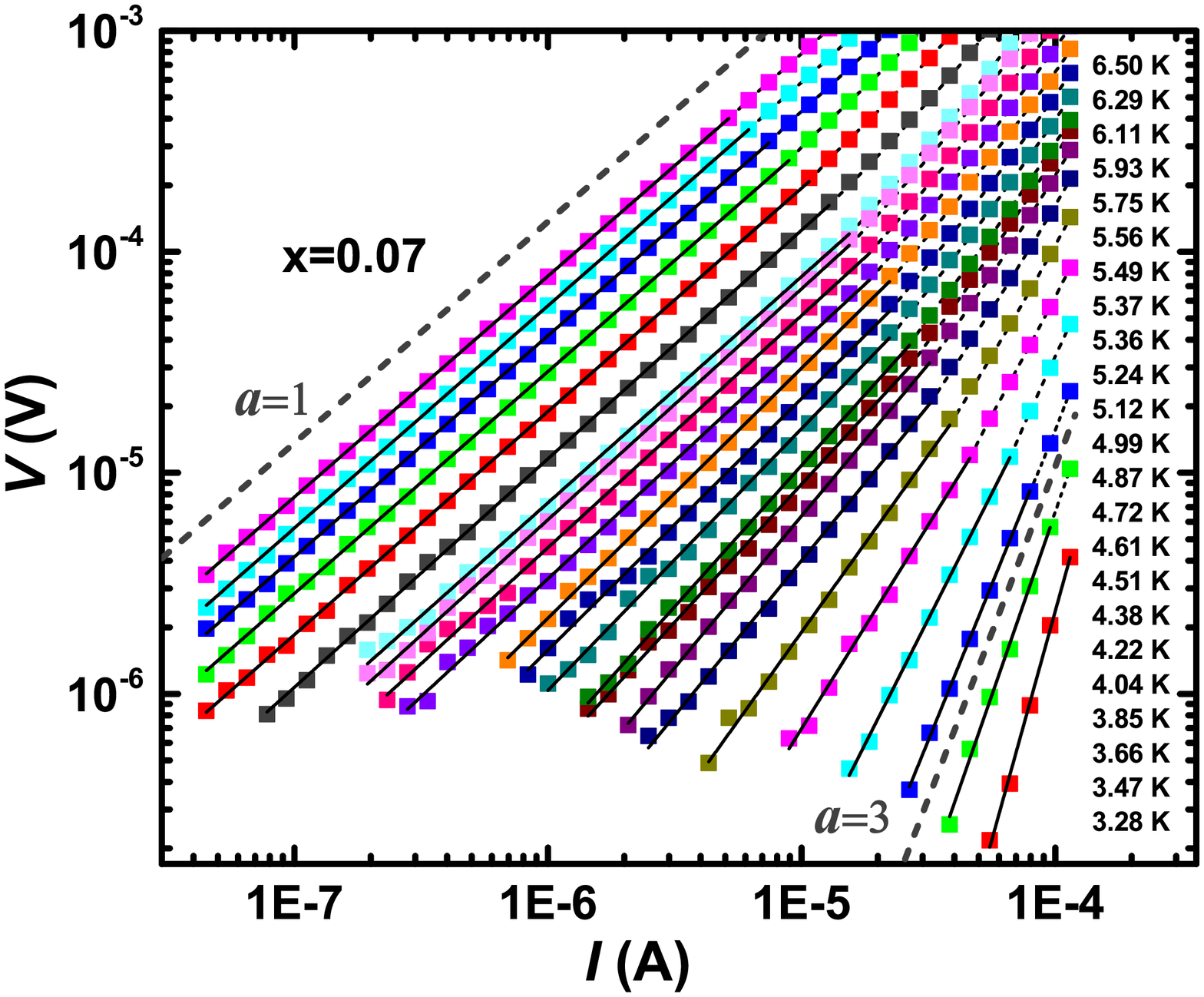}\llap{\parbox[b]{4.5in}{\textbf{(a)}\\\rule{0ex}{2.1in}}}
\includegraphics[width=8.5cm]{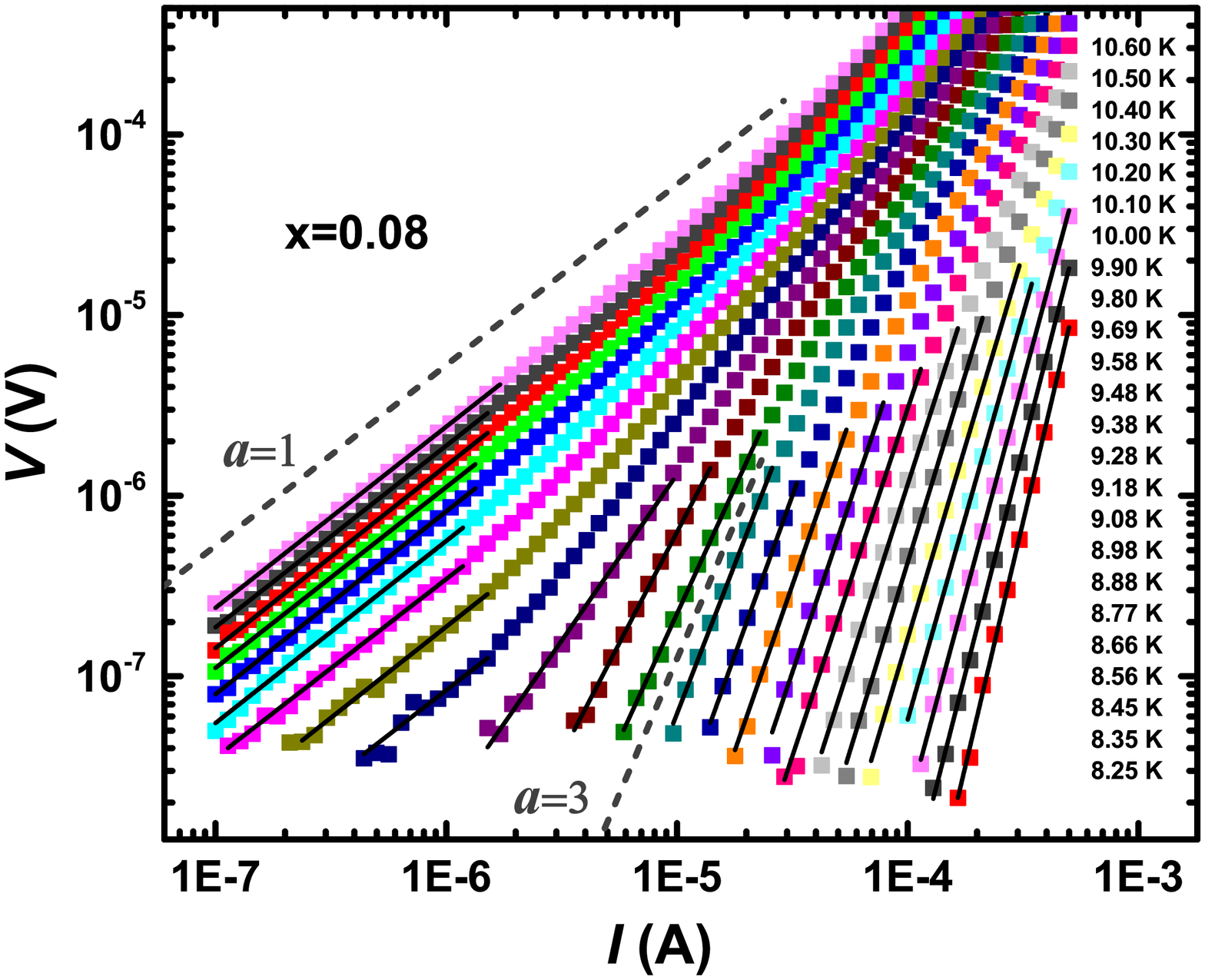}\llap{\parbox[b]{4.5in}{\textbf{(b)}\\\rule{0ex}{2.1in}}}
\includegraphics[width=8.5cm]{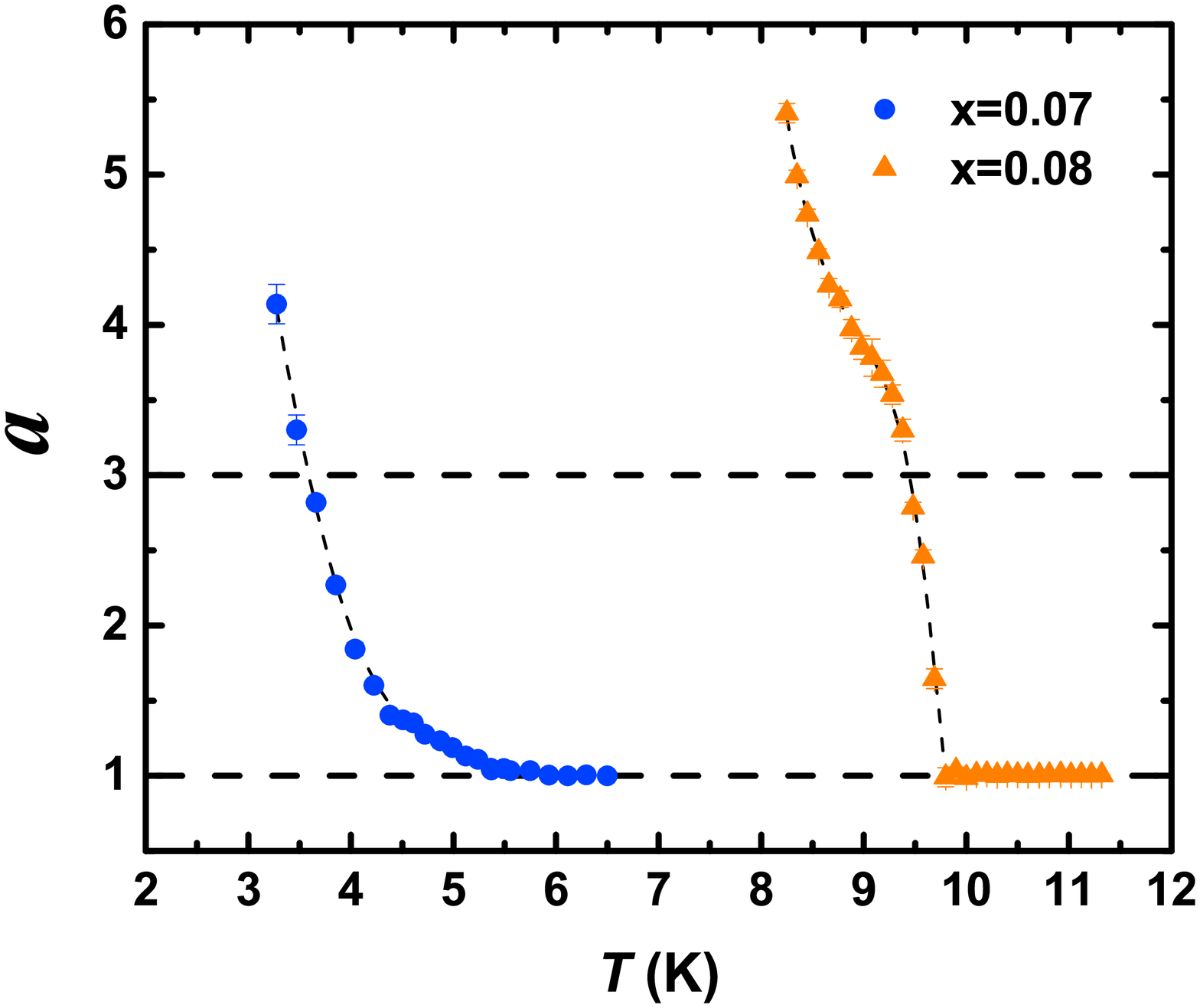}\llap{\parbox[b]{4.5in}{\textbf{(c)}\\\rule{0ex}{2.1in}}}
\caption{(a), (b) Voltage-current characteristics 
on a log-log scale for $x=0.07$ and $x=0.08$ samples, respectively, at different $T$, as shown.  At the lowest excitations, $V\propto I^{a(T)}$, where the solid lines are linear fits with the slopes corresponding to $a(T)$.  In both panels, the dashed lines with slopes $a=1$ and $a=3$ guide the eye.  (c) $a(T)$ for both samples.  The dashed line $a=3$ crosses the data at $(3.6 \pm 0.1)$~K and $(9.4 \pm 0.1)$~K for samples $x=0.07$ and $x=0.08$, respectively.}\label{fig:IV}
\end{figure}

The voltage-current characteristics are shown in Figs.~\ref{fig:IV}(a) and \ref{fig:IV}(b) on a log-log 
%
%%%%%%%%%%%% Figure 3 %%%%%%%%%%%%%%%
%
%%%%%%%%%%%%%%%%%%%%%%%%%%%%%%%
%
scale for the $x=0.07$ and $x=0.08$ films, respectively.  The power-law behavior $V\propto I^{a(T)}$ is observed at all $T$ in the low-current limit.  In that regime, the $V(I)$ dependence is thought to arise from the thermally dissociated vortex-antivortex pairs for $T>T_{BKT}$ and from current-induced dissociation for $T<T_{BKT}$.  At the highest currents in Figs.~\ref{fig:IV}(a) and \ref{fig:IV}(b), heating effects become important.  The temperature dependent exponents $a(T)$ were determined as the slopes of the linear fits of the data at the lowest currents [Figs.~\ref{fig:IV}(a) and \ref{fig:IV}(b)].   We note that, due to a large value of $t_c$, the fitting range, both in current and in temperature, is much wider than usual, \textit{i.e.} compared to systems that are clearly 2D, such as interfaces\cite{Reyren_2007} and films.\cite{Baturina_2012}  The values of $a(T)$ are presented in Fig.~\ref{fig:IV}(c) for both samples.  A steep change of $a$ from its Ohmic value ($a=1$) at high $T$ to large values $>3$ is indeed observed with decreasing $T$.  In particular, $a(T)$ in the $x=0.08$ sample exhibits a jump-like behavior as expected, but the $a(T)$ dependence is smoother in a more highly underdoped sample.  Nevertheless, $a$ reaches 3 at $T=(3.6 \pm 0.1)$~K and $(9.4 \pm 0.1)$~K for samples $x=0.07$ and $x=0.08$, respectively, close to their $T_{R=0}$ values and consistent with the assumption in Sec.~\ref{exp:para} that $T_{R=0}=T_{BKT}$.  

Even though the $T_{BKT}$ values will be determined more precisely in Sec.~\ref{th-J} by the theoretical analysis that takes into account the smearing of the BKT jump by the presence of inhomogeneities, we can estimate the order of magnitude of $J_s(T_{BKT})$ from the temperature where $a(T)=3$ using Eq.\ \pref{jump}. In the $x=0.07$ sample, for example, we have $J_s(T_{BKT}\approx 3.6$~K$)\approx2.3$ K.  Using this value and Eq.\ \pref{jbkt} expressed as\cite{yong_prb13}
\be
\label{defjn}
J_s[\mathrm{K}]=0.62 \, \frac{d_{BKT}[ {\angstrom} ] }{\lambda^2[\m \mathrm{m^2}]},
\ee
we find that, if the effective transverse length scale $d_{BKT}$ coincides with the film thickness ($\approx10^3$ \AA), $\lambda(T_{BKT})\approx 16~\m$m, while for $d_{BKT}\simeq d_c\approx 7$ \AA\, we obtain $\lambda(T_{BKT})\approx1.4~\m$m. Based on the doping and temperature dependences of the penetration depth measured in similar LSCO films,\cite{martinoli_prb96} we estimate that $\lambda(T_{BKT})$ does not exceed a value of 2-3 $\m$m for our $x=0.07$ sample.  Therefore, we find much better agreement between the results of our $I-V$ measurements and penetration depth studies by assuming that the  \textit{effective} sample thickness is somewhat larger than the interlayer spacing, but not as large as the whole thickness of the sample. As we shall see below, this conclusion is confirmed by a detailed comparison between $J_s(T)$ extracted from the $a(T)$ exponent and the theoretical prediction for the BKT behavior, when the non-trivial role of the vortex-core energy is taken into account. 

Finally, we remark that, in our samples, we do not expect to observe the Ohmic response in the $I-V$ characteristics caused by finite-size effects.\cite{weber_prb96,pierson_prb99,andersson_prb13,gurevich_prl08} Indeed, it is known that the dc $I-V$ curves probe the contribution of dissociated vortex-antivortex pairs separated by a distance $r^\ast=2\pi J_{s} cW/\Phi_0 I$.  Therefore, at small currents, which probe $r^\ast$ larger than the sample width ($W<L$),  the free vortices will dominate the resistance and the $I-V$ characteristics will be Ohmic.  On the other hand, the nonlinear behavior \pref{alpha} of the $I-V$ characteristics can only be seen when $r^\ast <W$, \textit{i.e.} above a threshold current\cite{review_minnaghen,benfatto_inho_prb09}  $I^{\ast}$,
\be
\label{istar}
I^{\ast}=\frac{2 J_s \pi c}{\Phi_0}\simeq \frac{4 k_B T_{BKT}\, c}{\Phi_0}. 
\ee
By using the above estimate $T_{BKT}\approx 3.6$ K for the $x=0.07$ sample, one gets $I^{\ast}\simeq 2.68 \times 10^{-8} \mathrm{(A/K)}\, T_{BKT}[\mathrm{K}] \sim 1\times 10^{-7}$~A.  In the presence of  inhomogeneous domains of size $L'<L$, the threshold current $I^*$ is expected\cite{benfatto_inho_prb09} to increase with respect to the estimate \pref{istar}.  However, since the  homogeneous  value \pref{istar} we found is considerably smaller than the currents at which the measurements are performed, finite-size effects are not expected to manifest themselves in our experiment. Indeed, 
Figs.~\ref{fig:IV}(a) and ~\ref{fig:IV}(b) show that, below $T_{BKT}$, the crossover from the nonlinear behavior \pref{alpha} back to the Ohmic one is not observed even at the lowest measured current.

\section{Theoretical Analysis of the Data}
\label{th}
\subsection{Superfluid stiffness}
\label{th-J}

We extract from Eq.\ \pref{alpha} the temperature dependence of the superfluid stiffness $J_s(T)$, which we analyze along the lines of the approach discussed earlier for both conventional\cite{benfatto_inho_prb09,mondal_bkt_prl11,yong_prb13,ganguli_prb15} and cuprate superconductors.\cite{benfatto_mu_prl07,benfatto_bilayer_prb08}  In Eq.\ \pref{alpha}, the temperature dependence of the superfluid stiffness $J_s(T)$ is due to both quasiparticle excitations, which induce a BCS-like suppression of $J^{BCS}(T)$ at all temperatures up to $T_c$, and vortex-like excitations, which become relevant near $T_{BKT}<T_c$. Since our $I-V$ measurements are rather close to $T_c$, we can assume for $J^{BCS}(T)$  a linear behavior, 
\be
\label{jbcs}
J^{BCS}(T)=J_0\left(\frac{T_c-T}{T_c}\right).
\ee
The effect of vortices is taken into account by solving the BKT renormalization-group equations, whose relevant variables are 
\bea
\label{defk}
K&=&\frac{\pi J^{BCS}(T)}{T},\\
\label{defg}
g&=&2\pi e^{-\beta\mu},
\eea
where $g$ is called the vortex fugacity ($\beta=1/k_{B}T$). Here $J^{BCS}(T)$ determines the value of $K$ at the shortest length scale of the problem, \textit{i.e.} the SC coherence length $\xi_0$, while the large-distance behavior will be determined by the presence or not of free-vortex excitations, described by the large-distance behavior of the vortex fugacity. The physical superfluid stiffness $J_s$ is then obtained by the numerical solution of the RG equations at large distances (see Appendix~\ref{app-RG} for more details). 

Apart from the starting value of $J^{BCS}(T)$, which can be determined by comparison with the data far from $T_{BKT}$, the second relevant energy scale in the problem is the ratio $\mu/J_s$. Here we take it as a free parameter, to be determined by the fit to the experimental data. This has to be contrasted to the usual $XY$-model description of the BKT transition, where $\mu/J_s$ is constrained to the value 
\be
\label{def_muxy}
\mu_{XY}=\frac{\pi^2}{2} J_s.
\ee
In general, the value of $\mu/J_s$ determines the temperature scale where significant deviations of $J_s$ from the BCS temperature dependence $J^{BCS}(T)$ start to become visible. Indeed, even though free vortices only start to proliferate at $T_{BKT}$, if a significant density of vortex-antivortex pairs already exists below $T_{BKT}$, it can renormalize (\textit{i.e}. suppress) the large-distance superfluid stiffness $J_s(T)$ with respect to its BCS behavior counterpart much before the BKT transition. In thin films of conventional superconductors it has been shown that this is the case.\cite{mondal_bkt_prl11,yong_prb13} Here $\mu/J_s\simeq 1$, as expected in ordinary BCS superconductors, and the measured $J_s(T)$ deviates from the BCS behavior significantly before 
the universal jump \pref{jump} occurs.   In contrast, it has been argued\cite{benfatto_mu_prl07,benfatto_bilayer_prb08} that, in cuprate superconductors, $\mu/J_s$ can even exceed the (large) value $\approx 4.9$ in Eq.~\pref{def_muxy}, where $J_s$ is now the stiffness of a single layer (\textit{i.e.} with $d_{BKT}=d_c$ in Eq.\ \pref{jbkt}). As we shall see below, this has relevant consequences for the determination of the effective transverse scale $d_{BKT}$ for the BKT transition in a bulk material or in a thick film, as it is in our case.

\begin{figure}[t]\centering{ 
\includegraphics[width=8cm]{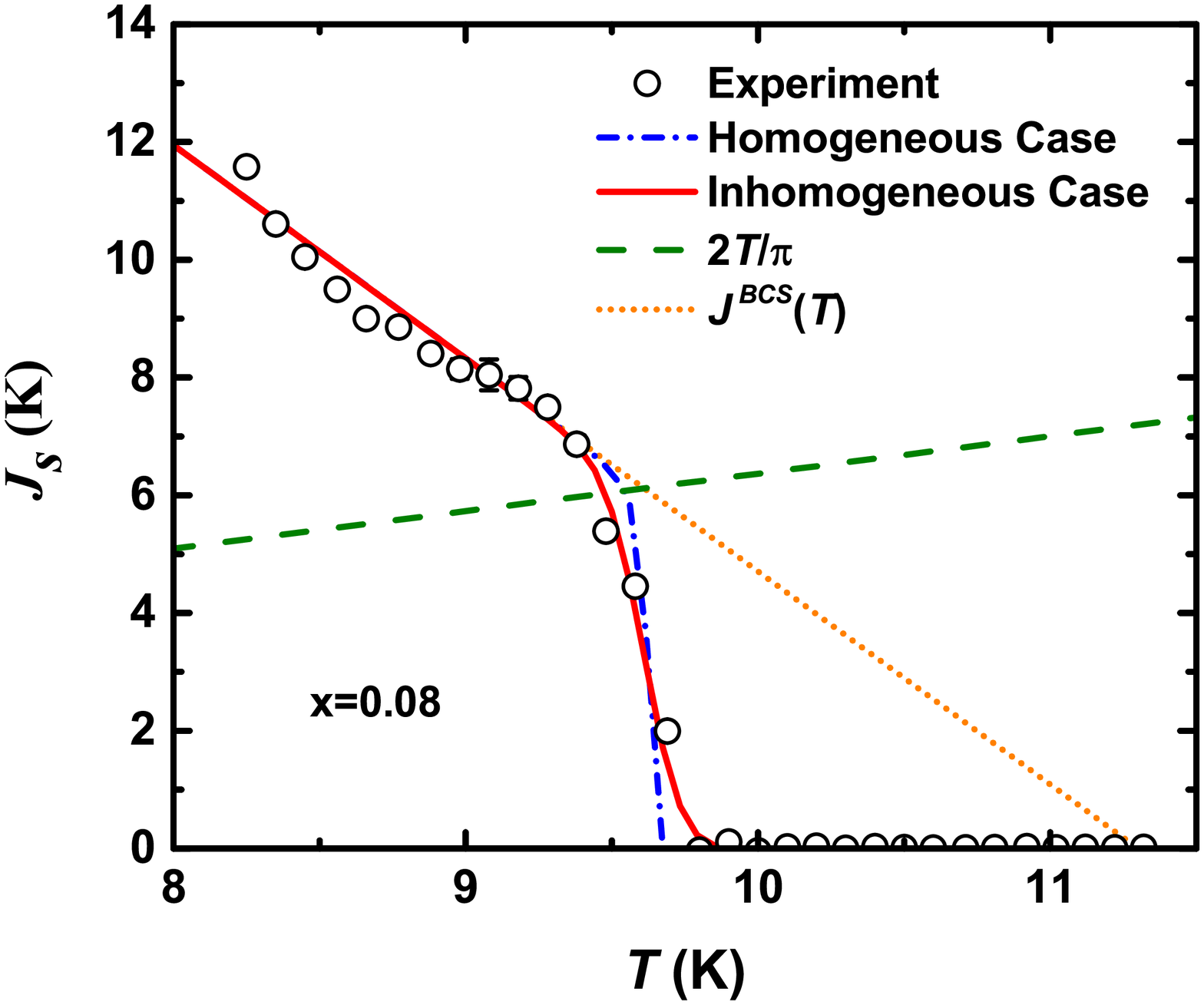}\llap{\parbox[b]{0.9in}{\textbf{(a)}\\\rule{0ex}{0.6in}}}
\includegraphics[width=8cm]{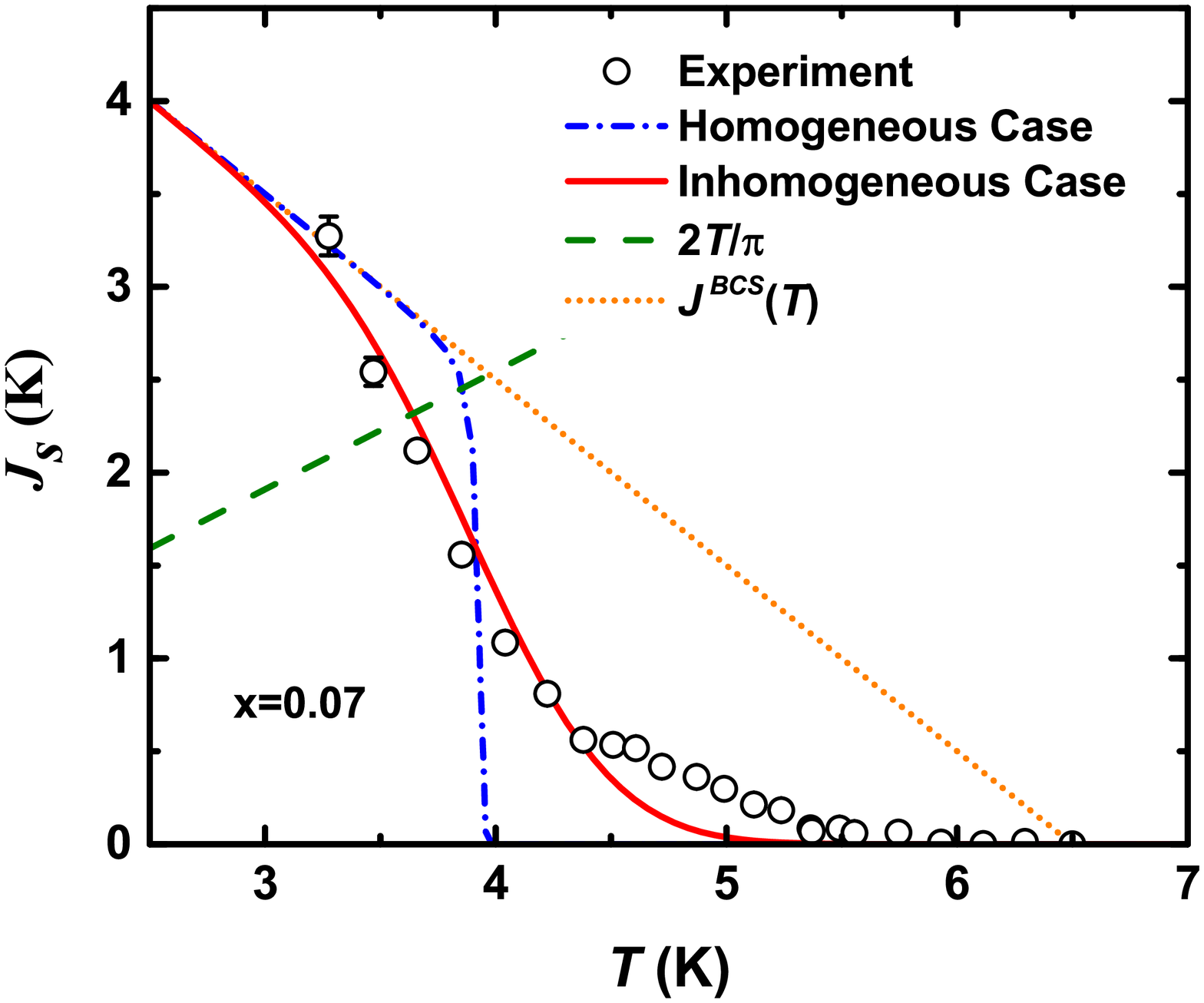}\llap{\parbox[b]{0.9in}{\textbf{(b)}\\\rule{0ex}{0.7in}}}}
\caption{Temperature dependence of the superfluid stiffness for the $x=0.08$ (a) and $x=0.07$ (b) samples: Comparison between the experiment and the theory, as described in the main text. The fitting parameters are listed in Table \ref{t-table}.}
\label{fig-Js}
\end{figure}

A second effect to be taken into account in the analysis of the experiments is the presence of inhomogeneity of the local SC properties, which have been clearly shown to be relevant in underdoped cuprates by means of STM analysis of underdoped samples.\cite{fischer_rmp07,yazdani_prl10}  Here we model\cite{benfatto_bilayer_prb08,mondal_bkt_prl11,yong_prb13} the presence of inhomogeneities by assuming that the local BKT critical temperature has a finite distribution about the most probable value, represented by the curve labeled ``Homogeneous'' in Fig.\ \ref{fig-Js}. The main effect of the inhomogeneity is then to smear out the universal jump \pref{jump}, the effect being larger for a wider probability distribution of the local $T_{BKT}$ values. More details are given in Appendix~\ref{app-RG}. 

The results for the two samples $x=0.07$ and $x=0.08$ are shown in Fig.\ \ref{fig-Js}, and the fitting parameters are summarized in Table \ref{t-table}. The BCS temperature dependence \pref{jbcs}, shown  in the figure with a dotted line, reproduces the data below the BKT 
%
%%%%%%%%% Figure 4 %%%%%%%%%%%%%%%
%
%
%%%%%%%%%%%%%%%%%%%%%%%%%%%%%
%
\begin{table}
\label{tablepar}
\begin{center}
\caption{Fitting parameters for Fig.\ \ref{fig-Js}.}
\label{t-table}
\begin{tabular}{|c|c|c|c|c|c|c|c|}
\hline 
doping &   $J_0$ (K) &$T_c$ (K) & $T_{BKT}$~(K) & $ \mu/J_s$  & $\delta/J_0$  &$t_c$ & $b_{theo}$\\
\hline
0.07 & 6.5 & 6.5 & 4 & 6.3 & 0.1 & 0.625 & 2.02\\
0.08 &  41 & 11.3 & 9.7 & 7 &0.01 &0.16 & 1.15\\
\hline 
\end{tabular}
\end{center}
\end{table}
transition very well, in particular for the $x=0.08$ sample where more experimental points are available. Here the sample inhomogeneity is very small (width of the distribution $\delta/J_0=0.01$; see also Appendix~\ref{app-RG}) and, accordingly, the homogeneous and inhomogeneous curves almost coincide, with a sharp downturn of $J_s(T)$ near $T_{BKT}$.  $T_{BKT}$ is defined here as the transition temperature for the homogeneous curve, which represents the most probable transition temperature for the sample.  We note that, since we also included the effects of the finite size of the system, which lead to some rounding of the $J_s(T)$ jump before $T_{BKT}$, even in the homogeneous case we do not observe a strictly discontinuous jump as in Eq.\ \pref{jump}, but $J_s$ vanishes continuously over a temperature range of a few mK.  For the $x=0.07$ sample, the inhomogeneity is larger ($\delta/J_0=0.1$), as expected for a  more underdoped sample, and this leads in particular to a longer superfluid tail above  $T_{BKT}$. In both samples, we extract a large value of the vortex-core energy, \textit{i.e.} $\mu/J_s=6-7$ or $\mu/\mu_{XY}\approx1.4$.  As explained above, this implies that the deviations of $J_s(T)$ from the BCS curve only occur near $T_{BKT}$. As a consequence, $T_{BKT}$ can be very well estimated by using the universal relation \pref{jump} with $J_s(T_{BKT}^-)$ replaced by $J^{BCS}(T_{BKT}^-)$, \textit{i.e.} 
\be
\label{tcapp}
J^{BCS}(T_{BKT})\simeq \frac{2T_{BKT}}{\pi}\Ra  \quad t_c\simeq \frac{2 T_c}{\pi J_0},
\ee
which is in very good agreement with the $t_c$ values listed in Table \ref{t-table}, obtained by the RG results.  It is apparent that the large separation between $T_c$ and $T_{BKT}$ in our samples is due to the presence of two concomitant effects in underdoped cuprate films: (i) the large mean-field critical temperature and (ii) the low superfluid stiffness, proportional to $J_0$ in Eq.\ \pref{tcapp}, due to correlations.\cite{emery_95,review_lee}  This has to be contrasted to conventional superconductors, where the BKT regime can only become visible when $J_0$ is suppressed by strong disorder, which also brings along unavoidable spurious effects connected to the inhomogeneity.\cite{mondal_bkt_prl11,yong_prb13,ganguli_prb15}  In addition, in systems like NbN, it has been shown that $\mu/J_s\simeq 1$, so the deviations of $J_s(T)$ from the BCS behavior occur much  {\em before} the intersection with the BKT line,\cite{mondal_bkt_prl11,yong_prb13} making the approximate estimate \pref{tcapp} much less reliable. 

Our finding of the large value of $\mu$ is an important result, since it confirms previous theoretical analysis\cite{benfatto_mu_prl07,benfatto_bilayer_prb08} in cuprates, and it allows us to understand the estimated value of $d_{BKT}\gtrsim d_c$ in our film, as discussed in Sec.~\ref{exp:IV}.  Since the measurements of $a(T)$ only access the areal superfluid stiffness \pref{jbkt} and thus do not allow for a separate determination of $d_{BKT}$ and $\lambda$, the comparison of the experimental data and the theory shown in Fig.\ \ref{fig-Js} has been done for the BKT transition in the pure 2D case.  On the other hand, we also know that our films are comprised of $\sim 10^2$ layers, with a weak interlayer Josephson coupling between them. In this case, it has been proven by previous theoretical work\cite{benfatto_mu_prl07,benfatto_bilayer_prb08} that, when $\mu/\mu_{XY}>1$, the BKT transition $T_{BKT}$ moves away from the value expected for a single, isolated layer $T_{BKT}^{n=1}$. In particular, according to the analysis of Refs.\ \onlinecite{benfatto_mu_prl07,benfatto_bilayer_prb08}, for the value of $\mu$ found above, one could expect that $T_{BKT}^n$ is about 30$\%$ larger than $T_{BKT}^{n=1}$, corresponding to $d_{BKT}\simeq (2-3)\, d_c$.  Indeed, by assuming $d_{BKT}\simeq 2\, d_c$ for the $x=0.07$ sample, for example, one can easily estimate $T_{BKT}=T_{BKT}^{n=2}\simeq 1.3\, T_{BKT}^{n=1}$ using the r.h.s. of Eq.\ \pref{tcapp}.  The value $d_{BKT}\simeq (2-3) \, d_c$ is consistent with the estimate based on the comparison to the penetration-depth measurements discussed in Sec.~\ref{exp:IV}. 

\subsection{Paraconductivity}
\label{th-para}

We note also that the value of $\mu$ extracted from the behavior of $J_s(T)$ is consistent with that obtained in Sec.~\ref{exp:para} from the analysis of the paraconductivity above $T_{BKT}$, even though the fits presented there do not include the effect of SC inhomogeneities. Indeed, we can show that for our samples the inhomogeneity has a relatively minor effect on the determination of the parameters entering the paraconductivity fit. To show this, we analyze the paraconductivity above $T_{BKT}$ by refining the analysis of Sec.~\ref{exp:para} with the inclusion of inhomogeneity. 

\begin{figure}
\centering
\includegraphics[width=8cm]{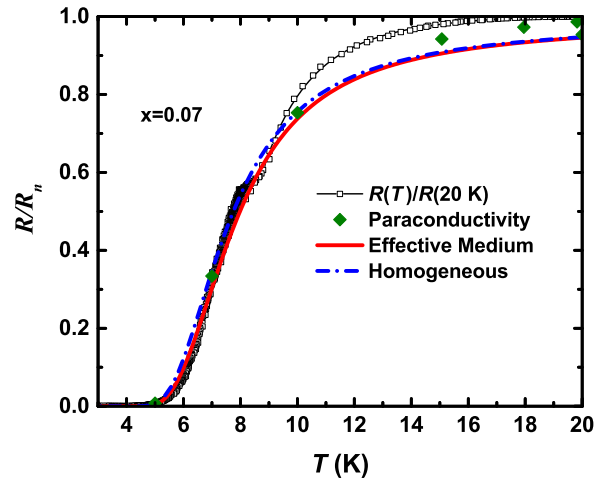}
\caption{Comparison between the $R(T)/R_n$ experimental data and the theoretical prediction obtained in the homogenous or inhomogeneous case. In the latter case, $R(T)$ is obtained as solution of a random-resistor-network problem in the effective-medium approximation, as explained in Appendix~\ref{app-RG}.  Solid diamonds: Paraconductivity was extracted from measurements in high magnetic fields (Sec.~\ref{exp:para}).}
\label{res}
\end{figure}

We can describe the measured resistivity as
\be
\label{int}
\frac{R}{R_n}=\left( 1+\frac{\D \s_{SCF}}{\s_n} \right)^{-1}=\left( {1+\left(\frac{2}{A}\sinh \frac{b}{\sqrt t}\right)^2 }\right)^{-1}.
\ee
In order to compare the theoretical predictions to a larger number of data points, in $\Delta\sigma_{SCF}(T)=\rho(T)^{-1}-\rho_{n}(T)^{-1}$, we approximate $\rho_n=1/\sigma_n$ with a constant, zero-field value measured at $T\gg T_{BKT}$.   Even though this procedure is less accurate far from $T_{BKT}$, this is not relevant for the discussion of the  effects near $T_{BKT}$, where the SCFs contribution diverges.  This is exemplified in Fig.\ \ref{res}, where paraconductivity obtained using this method is compared to that extracted from measurements in high magnetic fields (Sec.~\ref{exp:para}).  Indeed, the exact determination of $\rho_n$ becomes crucial only far from $T_c$, where $\Delta\sigma_{SCF}/\sigma_n$ becomes comparable to the difference ($\sim 10$\%) between its values obtained using those two methods.

For the $x=0.07$ sample, we choose $R_n\equiv R(T=20$~K), where the SCFs contribution to conductivity is only a few per cent [Fig.~\ref{fig:SCF}(a)].  
To account for the inhomogeneity, we follow the procedure discussed in  Ref.~\onlinecite{mondal_bkt_prl11}, and outlined in Appendix~\ref{app-RG}. We use the same distribution of local critical temperatures extracted from the analysis of $J_s(T)$ to generate a distribution of local resistivity values $R_i/R_n$ described by Eq.\ \pref{int} with the same local values $T_{BKT}^i,T_c^i$ computed above. Thus, only $A,b$ in Eq.\ \pref{int} are the fitting parameters. The global resistance of the sample is then determined by the corresponding random-resistor-network problem by means of the effective-medium approximation. Once again, to elucidate the role of inhomogeneity, we compare the results for the homogeneous and inhomogeneous case.  The ``Homogenous'' curve  in Fig.\ \ref{res} refers 
%
%%%%%%%%%%%%%% Figure 5 %%%%%%%%%%%%%%%
%
%%%%%%%%%%%%%%%%%%%%%%%%%%%
%
to the paraconductivity of a system with a single $T_{BKT}$ and $T_c$ realization, corresponding to the most probable value in the sample. Thus, this is the paraconductivity expected for a homogeneous system whose superfluid stiffness below $T_{BKT}$ is described by the ``Homogeneous'' $J_s(T)$ curve in Fig.\  \ref{fig-Js}(b). 

The results are shown in Fig.\ \ref{res} for the parameters $A=14$ and $b=2.55$, which are in good agreement with the results of the analysis in Sec.~\ref{exp:para}. The differences are due to the effect of the inhomogeneity, which is known\cite{caprara_prb11} to affect the slope of $R(T)$ 
above the transition. More importantly, $b=2.55$ is very close to the theoretical value $b_{theo}=2$ calculated from Eqs.\ \pref{best}, \textit{i.e.}
\be
\label{btheo}
b_{theo}=\frac{4}{\pi^2}\frac{\mu}{J_s}\sqrt{t_c},
\ee
by using the values of $\mu/J_s$ and $t_c$ extracted from the analysis of the $I-V$ characteristics \textit{below} $T_{BKT}$, and listed in Table \ref{t-table}.  The fit accurately reproduces the data up to $T\simeq 10$ K, which is an extremely large range of SCFs, similar to Fig.~\ref{fig:SCF}(a). However, the BKT fluctuation regime only extends up to $T_c\approx 6.5$~K and, above it, ordinary AL-like Gaussian fluctuations are at play. Finally, we note that some deviations start to occur above $T\simeq 10$ K. As we discuss in Appendix~\ref{app-RG}, this effect has already been observed in several families of cuprates,\cite{caprara_prb05,leridon_prb07} and it can be interpreted as a signature of a pseudogap state above $T_c$.\cite{caprara_prb05,varlamov_prb11}

\section{Discussion}

The analysis carried out in the previous Sections clearly demonstrates the occurrence of a BKT-like transition in our LSCO samples. This is confirmed both by the analysis of the paraconductivity above $T_{BKT}$ and by the analysis of the superfluid stiffness below $T_{BKT}$, as extracted from the $I-V$ measurements. Even though our films are thick, in the sense that $d$ is much larger than the SC coherence length, the possibility to see BKT physics is guaranteed by the layered nature of the system. As we discussed in Sec.~\ref{intro}, a weakly-coupled layered superconductor is an ideal candidate for observing the BKT physics,
since a layered structure ensures the best screening of the charged supercurrents.  Indeed, in this case the interaction between vortices in each plane is logarithmic up to a scale $\Lambda_J\simeq \xi_0/\sqrt{J_\perp/J_\pll}$ that grows as the stiffness anisotropy increases. While the behavior of $J_\perp/J_\pll$ as a function of doping in the LSCO family has not been systematically explored, in other cuprates it has been shown to decrease significantly with underdoping,\cite{hosseini_prl04} along with a general suppression\cite{martinoli_prb96,broun_prl07} of $J_\pll$ due to correlation effects.\cite{emery_95,review_lee} 
Under these conditions, one could expect to identify signatures reminiscent of the typical 2D BKT physics, such as an almost discontinuous suppression of the superfluid stiffness, even in a layered {\em bulk} sample.\cite{shenoy_prl94,friesen_prb95,benfatto_mu_prl07,sondhi_prb09}  On general grounds, the starting point of this reasoning is that, as demonstrated within several models\cite{shenoy_prl94,friesen_prb95,pierson_prb95,olsson_prb91,benfatto_mu_prl07,sondhi_prb09} the physics of a layered superconductor with a very weak interlayer coupling closely approaches that of an isolated 2D system. Indeed, even though the transition will ultimately have a 3D character, the 3D critical region is extremely reduced for weak interlayer Josephson coupling,\cite{shenoy_prl94,pierson_prb95} and it could even be masked in the experiments due to  finite-size effects or inhomogeneities of the type discussed in this manuscript.

Since in the BKT picture there exists a universal relation \pref{jump} between the transition temperature and the smallest superfluid stiffness beyond which vortex unbinding occurs, the idea that each layer is isolated  can lead to the naive expectation that $T_{BKT}$ is controlled by the stiffness of each isolated layer, \textit{i.e.} the value \pref{jbkt}  with $d_{BKT}=d_c$.  However, as predicted theoretically,\cite{benfatto_mu_prl07} this simple picture should be in part revised when the role of the vortex-core energy $\mu$, controlling the vortex fugacity $g\sim e^{-\beta \mu}$,  is taken into account. Indeed, in a layered BKT model the transition temperature is not controlled by the ``bare'' (\textit{i.e.} short-distance) values of $J_\perp/J_\pll$ and of the vortex density $g$, but by their large-distance behavior. Both energy scales grow at large distances, with opposite consequences: the increasing of $J_\perp/J_\pll$ tries to keep the system superconducting, while the increase of $g$ implies that vortices would like to proliferate making the system non-superconducting. While at some temperature $g$ will ultimately win, the counter-action of the interlayer coupling can move $T_{BKT}$ away from the temperature scale connected to the single-layer stiffness.  Thus, the effective stiffness to be used in Eq.\ \pref{jump} has to be computed from the definition \pref{jbkt} with a transverse length scale $d_{BKT}$ somewhat larger than $d_c$. In particular, as $\mu$ increases, the transition temperature moves farther away from the single-layer temperature scale.\cite{benfatto_mu_prl07}  The large value of the vortex-core energy obtained in our measurements suggests that, in strongly-underdoped LSCO samples, the relevant length scale controlling the BKT transition involves a few coupled layers, \textit{i.e.} $d_{BKT}\gtrsim d_c$.  This conclusion is in agreement with the estimate based on the measured superfluid stiffness \pref{jbkt}, \textit{i.e.} a combination $d_{BKT}/\lambda^2$, and the comparison to $\lambda$ measured in similar films.  

These findings, based on the analysis of the superfluid stiffness below $T_{BKT}$, are confirmed by the analysis of the paraconductivity above it. In particular, we have shown that the SC fluctuations above $T_{BKT}$ exhibit a BKT character near the transition, and then evolve into the ordinary Aslamazov-Larkin-type behavior expected for Gaussian (amplitude and phase) fluctuations. We fitted the data with the well-known Halperin-Nelson formula,\cite{HN} which interpolates between the two regimes, by constraining the fitting parameters according to the theoretical expectations for them.\cite{benfatto_inho_prb09} This procedure not only provides a consistency check of the validity of the BKT analysis, but it also allows us to confirm the estimate of the vortex-core energy extracted by the analysis of superfluid density. In agreement with previous findings in bulk cuprates,\cite{rullier_prl07,leridon_prb07,rullier_prb11} most of the fluctuation regime is dominated by Gaussian fluctuations with a marked 2D character, where the characteristic 2D unit is represented by a single layer, \textit{i.e.} $d_{AL}=d_c$. It is worth stressing that this result is  not in contradiction with the finding $d_{BKT}\gtrsim d_c$ for the BKT behavior. Indeed, in the case of Gaussian fluctuations, the dimensionality of the fluctuations is controlled only by the band-parameter anisotropy, \textit{i.e.} the ratio $t_\perp/t_\pll$ between interlayer and intralayer hopping, respectively. When this ratio is small, as it is in cuprates, one can see 2D fluctuations over a a broad temperature range.\cite{Varlamov_book} The crossover to 3D behavior, expected  in bulk materials, is here preceded by the vortex fluctuations, which drive the system towards a 2D BKT transition.  Even if the transition will ultimately have a 3D character, we do not identify the crossover to 3D fluctuations. This is consistent with the fact that the 3D critical regime, especially above the transition,\cite{pierson_prb95} is extremely reduced in a weakly-coupled layered system and, in addition, it gets masked by inhomogeneous effects that are mostly relevant at the transition. 

\section{Conclusions}

We have presented measurements of the in-plane transport properties of two strongly underdoped thick films of  \lasco.  Our results have (i) established the occurrence of a BKT-like transition and (ii) identified the typical transverse length scale that defines the equivalent two-dimensional unit controlling the BKT signatures in this layered system. 

The most striking signature of a vortex-driven phase transition emerges from the superfluid stiffness $J_s$, extracted from the exponent of the nonlinear $I-V$ characteristics across $T_{BKT}$. In both samples, we observe a rapid downturn of $J_s$ reminiscent of the well-known universal jump  expected in a 2D superconductor. A quantitative comparison with the theoretical predictions, which also include the effect of some unavoidable degree of inhomogeneity in the samples, strongly suggests a large energetic cost to create the vortex cores in the SC state. As a consequence, even though the interlayer Josephson coupling is weak, the vortex-pair unbinding occurs at a temperature larger than the one where each isolated layer would undergo the BKT transition.\cite{benfatto_mu_prl07} In other words, the characteristic energy scale controlling the BKT properties corresponds to the superfluid stiffness of a few layers. These results are confirmed by the analysis of the paraconductivity above $T_{BKT}$. Thanks to the few-K distance between the BKT ($T_{BKT}$) and mean-field ($T_c$) critical temperatures, we can clearly see that an initial BKT regime of fluctuations crosses over to an extended regime of 2D Aslamazov-Larkin-type Gaussian fluctuations. 

As we remarked above, the advantage of using highly underdoped thick films is that the intrinsically low value of the superfluid stiffness, due to the proximity to the Mott-insulating phase,\cite{emery_95,review_lee} allows us to achieve a large separation between $T_{BKT}$ and $T_c$ without simultaneously introducing a large disorder-driven inhomogeneity of the local SC properties. This has to be contrasted with the case of few-unit-cell thick films of cuprates,\cite{lemberger_natphys07,lemberger_prb12} which are usually much more sensitive to disorder, so that the BKT jump of the superfluid stiffness is usually lost with underdoping.\cite{lemberger_prb12} We note also that finite-frequency probes, such as the two-coil mutual inductance technique used in Ref.\ \onlinecite{lemberger_prb12}, can be potentially much more sensitive to disorder-induced inhomogeneity, as discussed recently within the context of films of conventional superconductors.\cite{ganguli_prb15}  In contrast, the superfluid density extracted from the $I-V$ characteristics is a purely dc probe,  and this can explain why we see a relatively sharp BKT jump even in our strongly underdoped samples. Whether these features are common to other families of cuprates is an interesting open question that certainly deserves further experimental and theoretical investigation.

\begin{acknowledgments}
We thank A. T. Bollinger and I. Bo\v{z}ovi\'{c} for the samples.  We acknowledge V. Dobrosavljevi\'c for useful discussions. High-field ($H>9$~T) measurements were performed in the National High Magnetic Field Laboratory (NHMFL) DC Field Facility.  This work was partially supported by NSF Grants No. DMR-0905843, No. DMR-1307075, and the NHMFL through the NSF Cooperative Agreement No. DMR-1157490 and the State of Florida.  L.B. acknowledges financial support by MIUR under projects FIRB-HybridNanoDev-RBFR1236VV, PRIN-RIDEIRON-2012X3YFZ2 and Premiali-2012 ABNANOTECH.
\end{acknowledgments}

\appendix

\section{Magnetoresistance}
\label{app-MR}

The $H^2$ dependence of the magnetoresistance is clearly observed at the highest $T$ and $H$ [Fig.~\ref{fig:MR}(b)].  As the temperature is lowered and SCFs become stronger, the $H^2$ region gets pushed to higher fields and the curvature of the $\rho(H)$ dependence at high $H$, in the normal state, becomes less obvious.  The same kind of behavior has been observed in other cuprates, \textit{e.g.} in \ybco \,(Ref.~\onlinecite{rullier_prl07})  and in overdoped,\cite{Vanacken2004, Cooper2009, Rourke2011} underdoped,\cite{Vanacken2004} and even non-superconducting\cite{Vanacken2004, Shi2013} LSCO very close to the onset of superconductivity.  

In underdoped LSCO, it is well known\cite{Ando_prl95, Marta_prl96, Greg_prl96, Vanacken2004} that the resistivity at high $H$ increases with decreasing $T$ (\textit{i.e. }$d\rho/dT < 0$), as seen also in Fig.~\ref{fig:MR}(a), reflecting the tendency towards an insulating ground state at high fields.\cite{Shi2014}  Nevertheless, deviations from Eq.\ \pref{mr-orb} still provide a good estimate of $H_{c}'(T)$, as discussed below.  While the precise reason for the applicability of Eq.\ \pref{mr-orb} remains an open problem beyond the scope of this study, we note that, in the regime of interest, the system remains in the (poor) metallic regime, as $k_{F} l\gtrsim 1$ ($k_F$ -- Fermi wave vector, $l$ -- mean free path), \textit{i.e.} the resistance per square per CuO$_2$ layer $\lesssim h/e^2$.

The quadratic dependence $H_{c}'(T)=H_{c}'(0)[1-(T/T_2)^2]$ (see Fig.~\ref{fig:SCF}(a) inset) was found also in \ybco\, (Refs.~\onlinecite{rullier_prl07, rullier_prb11}) and overdoped LSCO (Ref.~\onlinecite{Rourke2011}) giving us further confidence that the values of $H_{c}'(T)$ are reliable.  Furthermore, the $H=0$ onset temperature for SCFs, $T_2=29$~K, is consistent with the results from terahertz spectroscopy\cite{Bilbro2011} obtained on similar films, and those determined from the onset of diamagnetism\cite{Li_prb10} and the Nernst effect\cite{Wang_prb01} in LSCO crystals with similar $\rho(T)$ and $T_c$ values.  
We also find that $\mu_{0}H_{c}'(0)\simeq15$~T is in agreement with the value of the upper critical field obtained from specific-heat measurements\cite{Wang_epl08} on LSCO with a similar $T_c$.  Therefore, even though the magnetoresistance method that we employed to determine $H_{c}'$ and $\rho_n(H)$ has an inherent limitation in accuracy, we conclude that both the magnitude and the temperature dependence of the onset fields $H_{c}'$ are fairly consistent with those from other types of studies.  This consistency check confirms further that the observed onset of the $H^2$ magnetoresistance corresponds to the return to the normal state. 

\section{Renormalization-group analysis of the BKT transition for an inhomogeneous system}
\label{app-RG}

The BKT RG equations describe the large-distance behavior of the dimensionless quantities $K$ and $g$ introduced in Eqs.\ \pref{defk}-\pref{defg} above. They are given by\cite{bkt2,review_minnaghen,benfatto_review13}
\bea
\label{eqk} 
\frac{dK}{d\ell}&=&-K^2g^2,\\ 
\label{eqg} 
\frac{dg}{d\ell}&=&(2-K)g,
\eea
where $\ell=\ln r/\xi_0$ is the rescaled length scale with respect to the short-distance cut-off for the problem, represented by the SC coherence length $\xi_0$. The initial values of $K$ and $g$ are determined by the BCS value of the superfluid stiffness, Eq.\  \pref{defk}, which includes only the temperature dependence due to quasiparticle excitations. The effect of vortices is accounted by the RG flow at large distances, so that the physical superfluid stiffness \pref{jbkt} is identified with the limiting value of $K$ as one goes to large distances:\cite{nelson_prl77}
\be
\label{jlim}
J_s\equiv \frac{T K(\ell\ra \infty)}{\pi}.
\ee
The basic idea of the RG equations is to look at the competition at large scales between the superfluid stiffness and the vortex fugacity. When $g\ra 0$, it means that single-vortex excitations are ruled out from the system, which then remains superconducting. Indeed, as one can see from Eqs.\ \pref{eqk} and \pref{eqg}, when $g\ra 0$, $K$ goes to a constant and then $J_s$ from Eq.\ \pref{jlim} is finite. If instead $g\ra \infty$ at large distances, it means that vortices proliferate and drive the transition to the non-SC state, since $K\ra 0$. The large-scale behavior depends on the initial values of the coupling constants $K,g$, which in turn depend on the temperature. The BKT transition temperature is defined as the highest value of $T$ such that $K$ flows to a finite value, so that $J_s$ is finite. This occurs at the fixed point $K=2,g=0$, so that at the transition one always has $K(\ell=\infty)=2$, corresponding to the universal relation \pref{jump} quoted above.  By numerically solving Eqs.\ \pref{eqk}-\pref{eqg} at each temperature, while taking $T_c$ and $\mu/J_{BCS}$ as free parameters in the initial value, we obtain the curve labeled as ``Homogeneous'' in Fig.\ \ref{fig-Js}, with the parameters reported in Table \ref{t-table}.

To account for the presence of inhomogeneities, we follow the procedure discussed in previous publications for both conventional\cite{mondal_bkt_prl11} and cuprate\cite{benfatto_bilayer_prb08} superconductors.  We assume that the BCS superfluid density is described by Eq.\ \pref{jbcs} with the initial value $J_0^i$ randomly distributed according to a probability density $P(J_0^i)$ that we take, for instance, as Gaussian:
\be
\label{gauss}
P(J_0^i)=\frac{1}{\sqrt{2\pi}\delta}
\exp\left[-(J_0^i-J_0)^2/2\delta^2\right].
\ee
In the homogeneous case, the Gaussian distribution has zero width and only the value $J_0$ is allowed. In this case, one obtains the $J_s(T)$ curve labeled as ``Homogeneous'' in Fig.\ \ref{fig-Js}, and the corresponding $T_c$, $T_{BKT}$ are the ones reported in Table \ref{t-table}.  As we remarked in the text, we also add finite-size effects, by stopping the RG flow at the scale $\ell_{max}=L/\xi_0$ of the system size. As a consequence, even for the homogeneous case $J_s(T)$ does not display a real jump, but an extremely rapid downturn occurring over a few-mK temperature range.  In the inhomogeneous case, for each $J_0^i$ value distributed according to Eq.\ \pref{gauss},  we rescale the corresponding BCS temperatures as $J_0^i/T_c^i=J_0/T_{c}$ and we compute $J_s^i(T)$  and the corresponding BKT temperatures $T_{BKT}^i$ by the numerical solution of the RG equations \pref{defk}-\pref{defg} above. After obtaining
this set of $J_s^i(T)$ curves, we compute the sample stiffness as the  average one $J_{av}$, defined as
\be
\label{jav}
J_{av}(T)= \sum_i P(J_0^i) J^i_s(T).
\ee
When all the stiffness values $J_s^i(T)$ are different from zero, as it is the case at low temperatures, the average stiffness will be centered around the center of the Gaussian distribution \pref{gauss}, so that it will coincide with $J^{BCS}(T)$. However, by approaching $T_{BKT}$ defined by the average $J^{BCS}(T)$, not all the patches make the transition at the same temperature, so that the BKT jump is rounded and $J_{av}$ remains finite above the average $T_{BKT}$, in agreement with the experiments. In this analysis, we then have a second free parameter that is the width $\delta/J_0$ of the Gaussian distribution \pref{gauss}, However, all four parameters of the fit (average $J_0$ and $T_c$, ratio $\mu/J_s$ and $\delta/J_s$) affect in a rather independent way the shape of the overall stiffness. Indeed, $J_0$ and $T_c$ are essentially determined by the slope of the stiffness before the BKT transition, $\mu/J_s$ determines the location of the universal jump, whose smearing is controlled by $\delta$. Thus, even though some flexibility is possible in the values of the parameters listed in Table 
\ref{t-table}, these variations are  expected to be within $10\%-20\%$ of the quoted values. 

The inhomogeneity also influences the paraconductivity above $T_{BKT}$. To show this, we proceed in analogy with Ref.~\onlinecite{mondal_bkt_prl11} by mapping the spatial inhomogeneity of the sample in a random-resistor-network problem. In particular, we associate to each patch of initial stiffness value $J_0^i$ a normalized resistance $\rho_i=R_i/R_n$ obtained from Eq.\ \pref{int} by using the corresponding local values of $T_c^i$ and $T_{BKT}^i$ computed as outlined above.  The overall sample normalized resistance $\rho=R/R_n$ is then calculated in the so-called effective-medium-theory (EMT) approximation,\cite{sema} where $\rho$ is the solution of the self-consistent equation
\be
\label{semares}
\sum_i \frac{P_i (\rho-\rho_i)}{\rho+\rho_i}=0.
\ee
Here  $P_i$ is the occurrence probability of each resistor, which coincides with the distribution function \pref{gauss} of the local $J_0^i$ value used to compute the corresponding $\rho_i(T)$. The resulting $R(T)=\rho R_n$ is shown in Fig.\ \ref{res}, and it is compared to the one of the homogeneous case, \textit{i.e.} the $R(T)$ curve obtained when only the most-probable $J_0$ value of the distribution \pref{gauss} is realized. 

\begin{figure}[htb] 
\centering
\includegraphics[width=8cm]{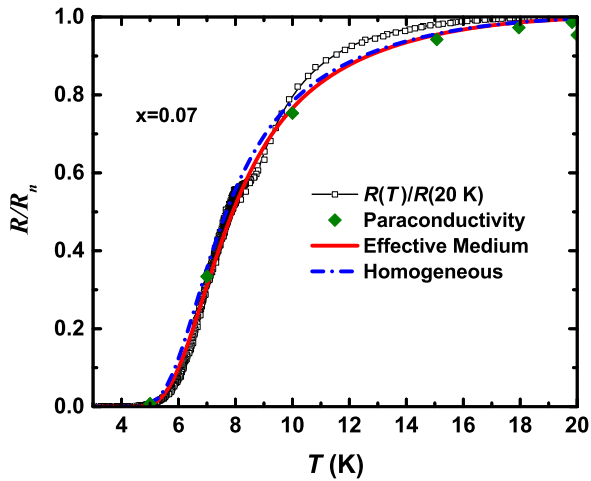}
\caption{Comparison between the $R(T)/R_n$ experimental data and the theoretical prediction obtained in the homogenous or inhomogeneous case for the modified HN correlation length \pref{xhn-corr}, including phenomenologically the pseudogap effect in the AL fluctuations.  Solid diamonds: Paraconductivity extracted from measurements in high magnetic fields (Sec.~\ref{exp:para}).}
\label{respseudo}
\end{figure}

As we observed in Sec.\ \ref{th}, for $T\gtrsim 10$ K the experimental paraconductivity saturates more rapidly than what is predicted by the HN interpolating formula. Since in this regime we are already exploring Gaussian fluctuations, such a failure is not correlated with the BKT character of the fluctuations, but it  pertains instead to the regime of ordinary Cooper-pairs fluctuations. Interestingly, such behavior has been already observed in several families of cuprates,\cite{caprara_prb05,leridon_prb07}  and it has been interpreted theoretically\cite{caprara_prb05,varlamov_prb11}  as an effect of the pseudogap. Indeed, by phenomenological  modelling of the suppression in the electronic density of states characteristic of a preformed pseudogap, one can reproduce a faster decay of the Cooper-pairs correlation length $\xi(T)$ in Eq.\ \pref{para} with respect to the standard AL prediction. Even though a detailed analysis of this issue is beyond the scope of the present manuscript, we nonetheless observed that a similar effect seems to be at play also in the case of our sample. To account for it within the HN interpolation scheme, we can for instance multiply the correlation length entering the paraconductivity formula \pref{para} by a function suppressing it around a temperature $T^*$ larger than $T_c$, 
such as
\be
\label{xhn-corr}
\xi(T)=\frac{2}{A}\sinh\frac{b}{\sqrt{t}}\exp(-(T/T^*)^4).
\ee
By introducing this correction factor in each normalized resistivity $\rho_i$ appearing in Eq.\ \pref{semares}, we obtain the (homogeneous and inhomogeneous) curves displayed in Fig.\ \ref{respseudo}. Here we used $T^*=19$ K, that is, approximately the temperature where magnetoresistance saturates. As one can see in Fig.\ \ref{respseudo}, our scheme now gives  an excellent agreement with the experimental data up to $T\simeq 20$ K.  In this high-temperature regime, the deviations of $R(T)$ from the magnetic-field extracted paraconductivity (symbols in Figs.\ \ref{res} and \ref{respseudo}) become sizeable, and the expression \pref{xhn-corr} reproduces the latter points well. 

%%%%%%%%%%%%%%% Figure 6 %%%%%%%%%%%%%%
%
%%%%%%%%%%%%%%%%%%%%%%%%%%%%%%%%%

\end{document}